\documentclass[a4paper,11pt]{article}

\usepackage{listings}             
\usepackage{csquotes}
\usepackage{xcolor}
\usepackage{comment}

\newcommand{\cei}{C_{\e 1}}
\newcommand{\ceii}{C_{\e 2}}

\def\karm{von  K\'arm\'an }
\def\D{\partial}
\def\a{\alpha}

\def\reich{{Reichardt's law}}
\def\wreich{{Reichardt's wall function}}
\def\ke{{k- \varepsilon}}
\def\e{\varepsilon}
\def\12{{\frac{1}{2}}}
\def\e{\varepsilon}

\def\ub{\bar{u}}
\def\vb{\bar{v}}
\def\wb{\bar{w}}

\def\V2{v^2}

\def\dx{\partial x}

\def\f13{{\frac{1}{3}}}
\def\23{{\frac{2}{3}}}
\def\32{{\frac{3}{2}}}

\def\D{\partial}

\def\sb{{\bar{s}}}

\usepackage{tikz,xifthen}
\usetikzlibrary{decorations.markings,arrows}
\usepackage{pgfplots}
\usepackage{color}
\usepackage{cancel}

\usepackage{subcaption}

\usepackage{graphicx,psfrag}
\usepackage{enumerate}
\usepackage{amsfonts}
\usepackage{latexsym}
\usepackage{amssymb}
\usepackage{times}
\usepackage{rotating}
\usepackage{pst-tree,pst-node}

\makeatletter
\pgfarrowsdeclare{myarrow}{myarrow}{
\pgfarrowsleftextend{-5\pgflinewidth}
\pgfarrowsrightextend{5\pgflinewidth}
}
{
            \newlength\pgf@my@length
            \pgf@my@length = 1cm
            \divide \pgf@my@length by 16
            \pgfpathmoveto{\pgfpoint{-.7\pgf@my@length}{0pt}}
            \pgfpathlineto{\pgfpoint{-.7\pgf@my@length}{-.8\pgf@my@length}}
            \pgfpathlineto{\pgfpoint{1\pgf@my@length}{0pt}}
            \pgfpathlineto{\pgfpoint{-.7\pgf@my@length}{.8\pgf@my@length}}
            \pgfpathlineto{\pgfpoint{-.7\pgf@my@length}{0pt}}
            \pgfusepathqfillstroke
}
\makeatother
\def\kdtree{\textbf{KDTree}}

\def\pycalc{\textbf{pyCALC-LES }}

\def\solidline{{$\textcolor{blue}{\overline{\hskip 0.5cm}}\ $}}
\def\solidredline{{$\textcolor{red}{\overline{\hskip 0.5cm}}\ $}}
\def\dashedline{{$\textcolor{red}{\overline{\hskip 0.1cm}\ \overline{\hskip 0.1cm}\ \overline{\hskip 0.1cm}\ }$}}
\def\dashedblueline{{$\textcolor{blue}{\overline{\hskip 0.1cm}\ \overline{\hskip 0.1cm}\ \overline{\hskip 0.1cm}\ }$}}
\def\dashedblackline{{$\textcolor{black}{\overline{\hskip 0.1cm}\ \overline{\hskip 0.1cm}\ \overline{\hskip 0.1cm}\ }$}}
\def\dashdottedline{$\overline{\hskip 0.1cm}\ \overline{\hskip 0.03cm}\ \overline{\hskip 0.1cm}\ $}

\def\solidline{\textcolor{blue}{\protect\rule[0.15\baselineskip]{0.6cm}{1.5pt}}\hspace*{0.1cm}}

\def\dashedline{\textcolor{red}{\protect\rule[0.15\baselineskip]{0.2cm}{1.5pt}}%
\hspace*{0.1cm}\textcolor{red}{\protect\rule[0.15\baselineskip]{0.2cm}{1.5pt}}\hspace*{0.1cm}}

\def\dashedblackline{\textcolor{black}{\protect\rule[0.15\baselineskip]{0.2cm}{2.5pt}}%
\hspace*{0.1cm}\textcolor{black}{\protect\rule[0.15\baselineskip]{0.2cm}{2.5pt}}\hspace*{0.1cm}}

\def\dashdottedline{\textcolor{black}{\protect\rule[0.15\baselineskip]{0.2cm}{1.5pt}}%
\hspace*{0.1cm}\textcolor{black}{\protect\rule[0.15\baselineskip]{0.05cm}{1.5pt}}%
\hspace*{0.1cm}\textcolor{black}{\protect\rule[0.15\baselineskip]{0.2cm}{1.5pt}}\hspace*{0.1cm}}

\usepackage[subfigure]{tocloft}

\usepackage[round]{natbib}
\usepackage[colorlinks=true,linkcolor=blue,anchorcolor=blue,urlcolor=blue,citecolor=blue]{hyperref}
\usepackage{lipsum}
\usepackage[left]{lineno}
\usepackage{hyperref}

\title{Hybrid LES/RANS for flows including separation: A new wall function using Machine Learning based on binary search trees} 
\author{
Lars Davidson \\ 
Div. of  Fluid Dynamics\\
 Mechanics and Maritime Sciences (M2)\\
Chalmers University of Technology, Gothenburg, Sweden \\
Email: lada@chalmers.se
}
\date{}

\begin{document}
\lstset{language=Python,
    numbers=right,
    stepnumber=1}

\maketitle

\begin{abstract}
Machine Learning (ML) is used for developing wall functions for Improved Delayed 
Detached Eddy Simulations (IDDES). The ML model is based on \kdtree\ which essentially is
a fast look-up table. It searches the nearest target datapoint(s) for which $y^+$ and $U^+$ 
are closest 
to the CFD $y^+$ and $U^+$. The target $y^+$ value gives the friction velocity.
Two target databases -- diffuser flow with opening angle $\a=15^o$ and hump flow -- are created from time-averaged  data
of wall-resolved IDDES (WR-IDDES, i.e. wall-adjacent cells at $y^+<1$).

The new ML wall function  is used to predict five test cases:
diffuser flow with opening angles $\a=15^o$ and  $\a=10^o$,
the hump flow, 
channel flow at $Re_\tau=16\, 000$ and flat-plate boundary layer.

A novel grid strategy is used. The wall-adjacent cells are large.  But further away from the wall, the
wall-normal cell distribution is identical to that of a WR-IDDES grid.
This new grid is found to improve the predictions compared to a standard
wall-function grid.

It is found that the number of cells for a wall-resolved IDDES grid (grid stretching $15\%$)
is a factor of $0.2\ln(Re_\tau)$ larger than
that of a standard wall-functions mesh (constant wall-normal grid cells). 

The new ML wall function is found to perform well compared to the WR-IDDES and better than the \wreich.

\end{abstract}


\section{Introduction}

Wall-resolved Large Eddy Simulation (LES) in which the wall-adjacent cell center  are located at $y^+ < 1$
 is affordable only at low Reynolds number ($x$, $y$, $z$ denote streamwise, wall-normal and spanwise direction, respectively;
$y^+ = u_\tau y/\nu$ where $u_\tau$ and $\nu$ denote friction velocity and kinematic viscosity, respectively).  At high
Reynolds number, LES can  be combined with a URANS treatment of the near-wall flow region. There are
different methods for combining LES and URANS  such as Detached Eddy Simulation
(DES)~\citep{spalart:jou:strelets:allmaras:97,spalart:00,shur:etal:08}, hybrid
LES/RANS \citep{davidson:peng:03,temmerman:hadiabdi:leschziner:hanjalic:05} and Scale-Adapted Simulations
(SAS)~\citep{menter:sas:theory:10,menter:sas:application:10}; for a review, see \cite{frohlich:terzi:08}.
The two first classes of models take the SGS length scale from the cell size whereas the last (SAS) involves
the \karm lengthscale.

In DES, the interface  between the LES region and the URANS
region  is supposed to be located in the outer part of the boundary layer.
However, it was found that in some flows -- for example, near the trailing edge of a airfoil -- it may occur that DES locates the interface
too close to the wall because the streamwise grid size is small. Hence 
the flow in the inner part of the boundary layer is
treated in LES mode with too coarse a  mesh. This results in a poorly resolved LES and inaccurate
predictions. 
Different proposals have been made to modify DES. The S-A DES model is modified by replacing the original DES lengthscale
$\tilde{d}$ with $\tilde{d}_{DDES}=d - f_d \max(0,d-C_{DES}\Delta)$~\citep{spalart:06} where $d$ is the wall distance, $C_{DES}$ is a
constant, $\Delta$ is the 
LES lengthscale and $f_d$ is a blending function which is zero
at the wall and one in the LES region.
In the $k-\omega$ SST DDES model ($k$ and $\omega$ denote modeled turbulent kinetic energy and specific dissipation, respectively), 
a shielding function is introduced which protects the boundary layer~\citep{menter:kunz:03,strelets:01}. 
IDDES is an \underline{I}mproved DDES~\citep{shur:etal:08}, see Section~\ref{IDDES}.

In the works cited above, URANS is used all the way to the wall (the wall-adjacent cells centres are located at $y^+ < 1$).
However, when making LES with wall functions (i.e wall modeled LES or wall stress models~\citep{larsson:12}), 
the wall-adjacent cells are located in the
fully turbulent region. 
The first work on  LES and  wall functions is that  of \cite{deardorff:70}.
The paper presents LES in fully-developed channel flow with periodic boundary conditions.
 
The wall distance  and the local velocity are usually input parameters in wall functions and they are  normally
taken at the wall-adjacent cells.  \cite{larsson:12,larsson:16} proposed to take these two quantities further away from the wall. The argument
is that the LES is not well-resolved at the wall-adjacent cells. 

It is interesting to compare the grid resolution requirements for different wall treatments.
 \cite{choi:12} find that the number of cells, $N$, for a wall-modeled LES (cubic cells) of a flat-plate boundary layer can be estimated
as $N \sim Re_L$ and for wall-resolved LES as $N \sim Re_L^{13/7}$ ($Re_L$ is the Reynolds number based on the free-stream velocity
and the length of the plate).
\cite{yang:21} reports slightly higher values, $N \sim Re_L^{1.14}$ and $N \sim Re_L^{2.72}$, respectively.
\cite{larsson:16} makes an estimate of the resolution requirement for wall-resolved LES of a NACA0012 at an angle of attack of $2.5^o$
at three different Reynolds numbers.
 They take into account
that the boundary-layer thickness varies along the surfaces. Their approximations are 
$N \sim Re_c^{0.4}$ and  $N \sim Re_c^{2}$ ($Re_c$ denotes the Reynolds number based on the free-stream velocity and the chord) 
for the outer and inner part of the boundary layer, respectively.

\cite{agrawal:24} investigates the grid-resolution requirements for separated flow. They define a length scale based on the
stream-wise pressure gradient $l_p =\nu/u_p$ where $u_p = ((\nu/\rho) dP/dx)^{1/3}$ ($P$ and  $\rho$ denote pressure and  density,
respectively). They report that in flows  which experience
 a stream-wise pressure gradient, the friction velocity is smaller than  $u_p$ (i.e.  $u_\tau < u_p$)
 which means that the 
timescale related to the pressure gradient is faster than 
the viscous timescale. Hence, the authors define a resolution requirement based on the stream-wise pressure gradient.
They find that the minimum resolution requirement is $\Delta/\min(l_p) \sim 10$, where $\Delta$ denotes the cell size.

The estimates of required grid resolution for wall-function meshes are given above. 
Now we will make an estimate of a hybrid LES/URANS mesh for which the
wall-adjacent cell centers are located at $y^+ < 1$. We set the wall parallel cell sides to $\Delta x = \Delta z = 1/N_0$ and
$\Delta y_{max} = 1/N_0$ (when making the estimates of required grids for 
wall-function meshes in \cite{choi:12}, $\Delta y$ was taken as constant and set to $1/N_0$). A geometric grid stretching, $\gamma$, is used
in the wall-normal direction but the stretching is stopped when $\Delta y = 1/N_0$. If the stretching is low, $\Delta y < 1/N_0$ for all
cells.  This occurs when $\gamma$ and $N_0$ satisfies the equation (see Appendix~\ref{appendix-wallf})
\begin{eqnarray}
\label{gamma-intro}
 Re_\tau (1-\gamma) = 1- \gamma^{\ln(Re_\tau/N_0)/\ln(\gamma)}
\end{eqnarray}
and when $\gamma$ is smaller.  $Re_\tau$ is the friction Reynolds number based on the friction velocity
and the boundary-layer thickness, $\delta$.
It is shown in  Appendix~\ref{appendix-wallf} that the ratio of the number of required cells in the wall normal-direction, $N_y$, for
a wall-stretched grid
to that for a wall-function grid is asymptotically
$\ln(Re_\tau)$ when Eq.~\ref{gamma-intro} is satisfied ($\gamma \simeq 1.04$)
and $0.2\ln(Re_\tau)$ for $\gamma =1.15$.

Hence it is found  that the reduction in CPU cost is rather small. Nevertheless, 
 \cite{herr:probst:21} reports a CPU reduction of 61\% when using wall functions compared to using WR-DES.
 Other advantages of using wall functions is that the viscous
restriction on the time step in compressible solvers is reduced. Furthermore, the ratio of the cell sides (i.e. the ratio 
of the streamwise cell side to the wall-normal one) is also reduced which means that the condition number of matrix, $M$, in the discretized
equations $MU = b$ ($U$ and $b$ are the solution vector (velocity components, pressure \ldots) and  the
right-hand side, respectively) is reduced.
Another advantage is that some works \citep{larsson:12,larsson:16,davidson:23-url,agrawal:24} 
report that the problem of log-layer mismatch (LLM) is eliminated when using wall functions.


There are not too many studies in the literature on ML for improving wall functions. 
\cite{zhou:21} use Neural Network to train an ML wall function. They make  wall-resolved DES (i.e. $y^+ < 1$ for the wall-adjacent cells) of periodic hill flow
at three different Reynolds numbers. They use instantaneous snapshots to train the ML wall function.  The output parameter is the wall-stress
vector and the input parameters are the wall-normal distance, the velocity vector and the pressure gradient. They use the ML
wall function for predicting channel flow at $Re_\tau  = 1\,000$ and  $Re_\tau  = 5\,200$ and they find that the ML wall function performs
very well. Then they apply the ML wall function  to the periodic  and they get very poor agreement with wall-resolved LES.
In~\cite{Tieghi:20} they
use a time-averaged high-fidelity IDDES simulation to train a neural network for improving the predicted
modeled turbulent kinetic used in wall functions (RANS). 
In \cite{ling:17} they use neural network to improve
the predicted wall pressure to be used in fluid-structure interactions. Their target is the wall pressure spectrum and the input
parameters are the pressure power spectra above the wall.
\cite{dominique:22} use neural network to predict the wall pressure spectra. Their input data are boundary-layer 
thicknesses (physical, displacement and momentum), streamwise pressure gradient and 
wall shear stress which are taken from experiments and high-fidelity DNS/LES in the literature.
In \cite{Bae:22} they use a neural network to create a pre-multiplication 
factor of the velocity-based wall model
(VWM) and a  log-law based wall model (LLWM). Then they introduce a reward factor, $r_n$,  at each time step $n$.
\cite{davidson:23-url} presents a new ML wall-function based on Support Vector Regression (SVR). The SVR is trained
using instantaneous $y^+$ and $U^+$ in fully developed channel flow at $Re_\tau  =5\, 200$. The ML model is
then used for predicting channel flow at $Re_\tau  =16\, 000$ and a flat-plate boundary layer. Good agreement with experimental
data is obtained.

A novel grid strategy is used in the present work. The wall-adjacent cells (cell number $j=0$) are large.
But further away from the wall (cell number $j >   0$), the wall-normal
cell distribution is identical to that of a WR-IDDES grid. 
The advantage of this grid strategy is that the gradients are much more accurately computed since the grid for cells $j >  0$
are much finer than for a standard wall-function grid.
This new grid is found to improve the predictions
compared to a standard wall-function grid.

\cite{gritskevich:17} made a comparison of IDDES with wall functions and WR-IDDES. They report that wall functions were essentially 
as accurate as WR-IDDES provided that the wall-normal grid spacing when using wall functions was smaller than one percent of the
boundary-layer thickness. Their finding supports the use of the novel grid strategy.

\cite{romanelli:23} use machine-learning for improving wall-functions for RANS (Reynolds-Averagared Navier-Stokes). 
They split their computational
grid (with wall-adjacent cell centers at $y^+ < 1$) 
into a near-wall region and an outer region. The separation line between the two regions is denoted the interface. 
No discretized equations are solved in the near-wall region, but the solution variables (velocity  componenents, pressure, \ldots)
in this region are set and used as ghost cells and serve as boundary conditions for the outer region. Incidentally, the resulting  grid strategy
in their work is the same as the novel grid strategy employed in the present study.

The ML model used in the present work is based on Python's \kdtree\ which essentially is a fast look-up table.
\kdtree\ computes the distance between the vectors $ \mathbf{X}_i$ (target database) and $\mathbf{x}_j$
(the wall-adjacent cells in the CFD simulation)
for all samples $i$ and $j$ and finds the $K$ smallest distances for each $j$.  The vectors $ \mathbf{X}_i$ and $\mathbf{x}_j$
have two columns, $y^+$ and $U^+$. The wall-shear stress for the wall-adjacent CFD cell $j$
is obtained by finding the closest cell (in terms of $y^+$ and $U^+$) in the target database.
The friction velocity is then obtained from  $y^+_{target}$
which is used 
for setting boundary conditions for  the wall-parallel velocity, $k$ and $\e$. 

The paper is organized as follows. The numerical method and turbulence model are presented in the next section.
Then we show how the two target databases are created using wall-resolved IDDES (WR-IDDES). In the next section we present the ML model
followed by a description of the \wreich. Then the results are presented and the paper ends with concluding remarks.

\section{Numerical method and turbulence model}

\label{sec:num}

The finite volume code \textbf{pyCALC-LES}~\citep{pyCALC-LES} is used. It is written in Python and is
fully vectorized (i.e. no \texttt{for} loops).
 The solution procedure is based on fractional step. Second-order central
differencing is used in space for the momentum equations and Crank-Nicolson is used in time.
For $k$ and $\e$, the hybrid central/upwind
scheme is used together with first-order fully-implicit time discretization. 
The code runs fully on the GPU. On a desktop, it runs approximately 50 times faster on the GPU (Nvidia RTX A6000) than on the CPU
(Intel i7-13700) for a  hump flow simulation using the present ML wall function.

\subsection{The governing equations}

The equations read
\begin{eqnarray}
\label{contLES}
\frac{\D \vb_i} {\D x_i}& =&  0\\
\label{LES}
\frac{\partial  \vb_i}{\partial t}  + 
\frac{\D \vb_i \vb_j} {\dx_j} &=&  
- \frac{1}{\rho} \frac{\partial \bar{p}}{\dx_i}  + \frac{\D}{\D x_j}\left[(\nu + \nu_{t}) \frac{\D \vb_i}{\D x_j}\right] 
\end{eqnarray}
where $\vb_i$, $\rho$, $p$, $\nu$, $\nu_{t}$ denote velocity vector, , density, pressure, kinematic viscosity
and kinematic, turbulent (i.e. modeled)  viscosity, respectively. Index $1$, $2$ and $3$ correspond to $x_1$, $x_2$ and $x_3$
respectively (i.e. $x$, $y$ and $z$).

\subsection{The underlying RANS turbulence model}

\label{AKN}

The AKN low-Reynolds number for IDDES (see Section~\ref{IDDES}) reads~\citep{abe:kondoh:94}
\begin{eqnarray} 
\label{rn-pans-ku-eu}
  \frac{\D k}{\D t} + \frac{\D \vb_j k}{\D x_j} &=&
\frac{\D}{\D x_j} \left[\left(\nu + \frac{\nu_t}{\sigma_{k}}\right)\frac{\D k}{\D x_j}\right] + P_k - \psi\e \\
\frac{\D \e}{\D t} + \frac{\D \vb_j \e}{\D x_j} & =& \frac{\D}{\D x_j} \left[\left(\nu +
\frac{\nu_t}{\sigma_{\e}}\right)\frac{\D \e}{\D x_j}\right] + C_{\e1} P_k\frac{\e}{k} -
C_{\e2}f_2\frac{\e^2}{k} \nonumber\\ 
\nu_t&=& C_\mu f_\mu \frac{k^2}{\e},  \quad
P_k = \nu_t \left(\frac{\D \vb_i}{\dx_j} + \frac{\D \vb_j}{\dx_i}\right)\frac{\D \vb_i}{\dx_j} \nonumber\\
C_{\e1} &=& 1.5, \quad C_{\e2} = 1.9, \quad C_\mu = 0.09, \quad \sigma_k = 1.4, \quad\sigma_\e = 1.4   \nonumber
\end{eqnarray}
where $k$ and $\e$ denote the modeled turbulent kinetic energy and its dissipation, respectively.
$\psi$ is defined in Eq.~\ref{psi-eq}.
The damping functions are defined as
 \begin{eqnarray*}
f_2 &=& \left[1 - \exp\left(-\frac{y^{\ast}}{3.1}\right)\right]^2 \left\{1 - 0.3
 \exp\left[- \left(\frac{R_t}{6.5}\right)^2\right]\right\}\\ 
 f_{\mu} &=& \left[1 -
 \exp\left(-\frac{y^{\ast}}{14}\right)\right]^2 \left\{1 + \frac{5}{R_t^{3/4}} \exp\left[-
 \left(\frac{R_t}{200}\right)^2\right]\right\}  \\
 y^* &=& \frac{u_\e d}{\nu},  \quad u_\e = \left(\e \nu)\right)^{1/4}, \quad Re_t  = \frac{k^2}{\nu \e}
 \end{eqnarray*}
where $d$ denotes the distance  to the wall.
The wall boundary condition is implemented by setting $\e$ at the wall-adjacent cells as
\begin{equation}
\label{chienBC}
\e =2\nu \frac{k}{d^2}
\end{equation}
When wall functions are used, the boundary conditions are set according to Eq.~\ref{bc}.

\subsection{The IDDES model}

\label{IDDES}

The Improved Delayed Detached Eddy Simulation method~\citep{shur:etal:08} is used when creating
the database as well as when using wall functions. 
The coefficient $\psi$ in Eq.~\ref{rn-pans-ku-eu} is computed as
\begin{eqnarray}
\label{psi-eq}
\psi =\frac{l_u}{L_{hyb}}, \quad l_u = \frac{k^{3/2}}{\e}
\end{eqnarray}
where $L_{hyb}$ is the usual IDDES length scale~\citep{shur:etal:08}. 
The IDDES is an improved DES model which is a hybrid RANS-LES method in which URANS
is used near the wall and LES is used further away from the wall. The IDDES includes two branches, WM-LES (wall-modeled LES)
and DDES (delayed DES). In WM-LES, the interface between URANS and LES is located in the logarithmic region whereas in DDES the entire
boundary layer is treated in URANS mode. In the present work, the WM-LES branch is used in all simulations.

For convenience, the procedure how to obtain 
the IDDES length scale is summarized below.
\begin{eqnarray}
\label{f_e}
L_{hyb}= f_d (1+ f_e) l_u + (1-f_d)l_c, \quad l_c =  C_{DES} \Delta
\end{eqnarray}
where the $\Delta$ length scale is defined as
\begin{eqnarray*}
\Delta = \min \left\{ \max \left[ C_w d_w , C_w h_{max} , h_{wn} \right] , h_{max} \right\}     
\end{eqnarray*}
and
$C_w = 0.15$, 
$d_w$ is the distance to the closest wall and
 $h_{wn}$ is the grid step in the wall normal direction.
The blending functions
$f_d$ and $f_e$ read
\begin{equation}
\label{eq:tifd}
f_d = \max \left\{ \left( 1 - f_{dt} \right) , f_B \right\}  
\end{equation}
\begin{equation}
\label{eq:fe}
f_e = \max{\left\{ \left( f_{e1} - 1 \right) , 0 \right\} } f_{e2}
\end{equation}
where the functions $f_{dt}$ and $f_B$ entering Eq.~\ref{eq:tifd} are given by
\begin{equation}
\label{eq:fdt}
f_{dt} = 1 - \tanh \left[ \left( 8 r_{dt} \right)^3 \right] 
\end{equation}
\begin{equation}
f_B = \min \left\{ 2 \exp \left( -9 \alpha^2 \right) , 1 \right\}  
\end{equation}
with
\begin{eqnarray*}
\alpha = 0.25 - d_w / h_{max} ~.
\end{eqnarray*}
The functions $f_{e1}$ and $f_{e2}$ in Eq.~\ref{eq:fe} read
\begin{eqnarray*}
f_{e1} = \left \{\begin{array}{ll} 2 \exp \left( -11.09 \alpha^2 \right) & {\rm if} ~ 
\alpha \ge 0 
 \\ 2 \exp \left( -9 \alpha^2 \right) & {\rm if} ~ \alpha < 0
 \end{array} \right .
\end{eqnarray*}
and
\begin{eqnarray*}
f_{e2} = 1 - \max \left\{ f_t , f_l \right\} 
\end{eqnarray*}
where the functions $f_t$ and $f_l$ are given by
\begin{eqnarray*} 
f_t = \tanh \left[ \left( c_t^2 r_{dt} \right)^3 \right] \label{eq:ft} \\
f_l = \tanh \left[ \left( c_l^2 r_{dl} \right)^{10} \right] \label{eq:fl}~. 
\end{eqnarray*}
The constants $c_t$ and $c_l$
 given the same values as in the $k - \omega$ SST model, i.e.   $c_t = 1.87$ and
$c_l = 5$~\citep{shur:etal:08}.
The quantities $r_{dt}$ (also entering Eq.~\ref{eq:fdt}) and $r_{dl}$, are defined as follows
\begin{eqnarray*} 
r_{dt} &=& \frac{\nu_t}{\kappa^2 d_w^2 \max \left\{ |\sb| , 10^{-10} \right\} } \nonumber   \\
r_{dl} &=& \frac{\nu}{\kappa^2 d_w^2 \max \left\{|\sb| , 10^{-10} \right\} }  
\end{eqnarray*}

\section{Creating the database}

\label{database}

\begin{figure}
\centering
\begin{subfigure}[c]{0.5\textwidth}
\begin{tikzpicture}[xscale=0.13, yscale=0.13]
\begin{axis}
[x=1cm,y=1cm,xmin=0,xmax=46,ymin=0,ymax=3.4,
axis x line=none,
axis y line=none
]
\addplot [mark=none,fill=gray]file {contour-iddes.txt};
\end{axis}
%
\draw [myarrow-myarrow] (0,-4.0) -- (6,-4.0);
\node [above] at (3,-4.0) {$L_1$};


\draw [myarrow-myarrow] (6,-4.0) -- (23.52,-4.0);
\node [above] at (14.8,-4.0) {$L_2$};

\draw [myarrow-myarrow] (23.52,-4.0) -- (46.00,-4.0);
\node [above] at (35,-4.0) {$L_3$};

\draw [myarrow-myarrow] (6,5.4) -- (6,2.4) -- (9,2.4);
\node [below right] at (8.3,3.3) {$x$};
\node at (6.0,6.4) {$y$};

\node [above] at (25,3.0) {slip wall};

\node [above] at (26,-3.1) {wall};

\node [left] at (0.0,2.9) {$h$};

\node [right] at (46.0,2.2) {$H$};

\draw [myarrow-myarrow] (18,0) -- (18,3.4);
\node [right] at (18.0,2.0) {$H_{max}$};

\node [right] at (11.9,1.1) {$\a$};
\end{tikzpicture}
\caption{Geometry.}
\label{diffuser-domain}
\end{subfigure}%
\begin{subfigure}[c]{0.5\textwidth}
\centering
\includegraphics[scale=0.3,clip=]{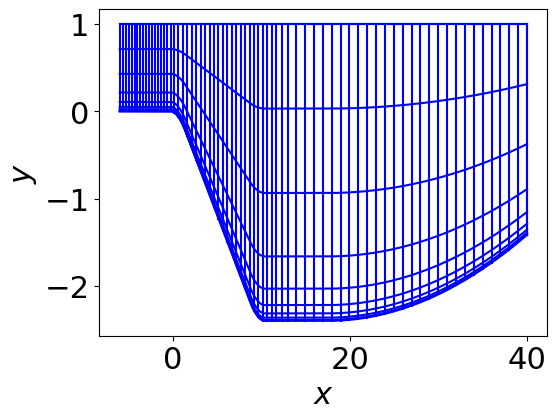}
\caption{Grid, $x-y$ plane (not to scale). $700\times 90$ cells. Every $10^{th}$ grid line is shown.}
\end{subfigure}
\caption{Diffuser, $\a=15^o$.}
\label{diff}
\end{figure}

\begin{figure}
\centering
\begin{subfigure}[t]{0.33\textwidth}
\centering
\includegraphics[scale=0.3,clip=]{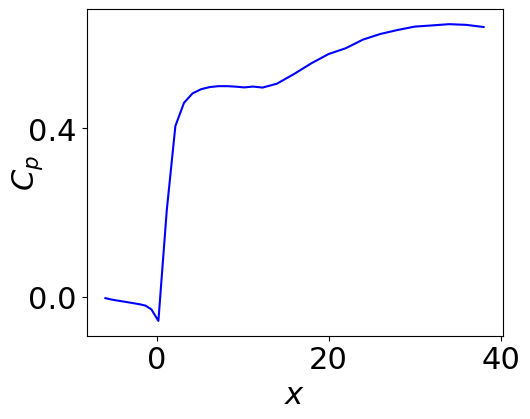}
\caption{Pressure coefficient.}
\end{subfigure}%
\begin{subfigure}[t]{0.33\textwidth}
\centering
\includegraphics[scale=0.3,clip=]{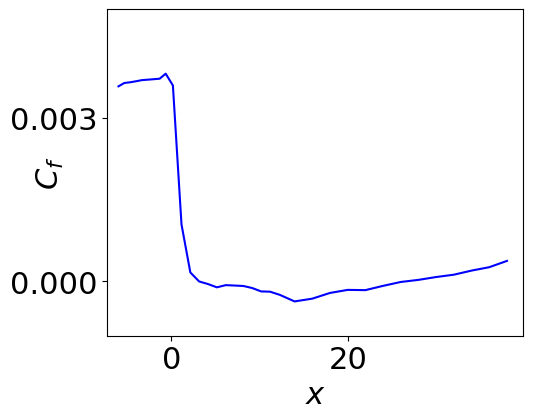}
\caption{Skin friction.}
\end{subfigure}
\begin{subfigure}[t]{0.33\textwidth}
\centering
\includegraphics[scale=0.3,clip=]{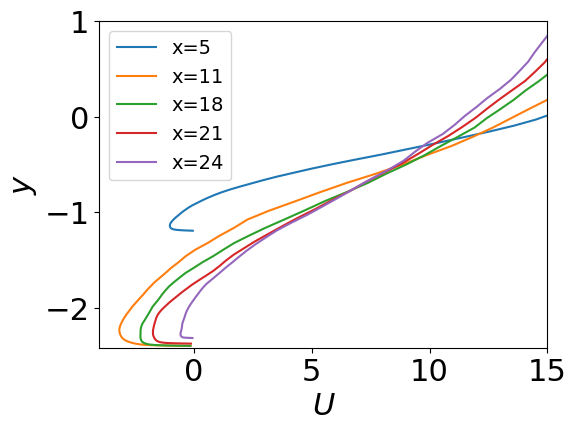}
\caption{Velocity.}
\end{subfigure}
\caption{Diffuser flow. Target database.}
\label{cp-cf-vel-diff}
\end{figure}

\begin{figure}
\centering
\begin{subfigure}[t]{0.5\textwidth}
\centering
\includegraphics[scale=0.30,clip=]{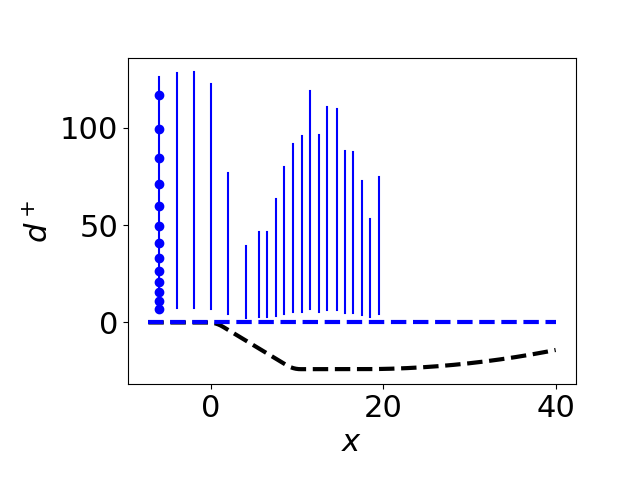}
\caption{Data points of $d^+$ vs. $x$.}
\label{target-x-yplus-diff}
\end{subfigure}%
\begin{subfigure}[t]{0.5\textwidth}
\centering
\includegraphics[scale=0.30,clip=]{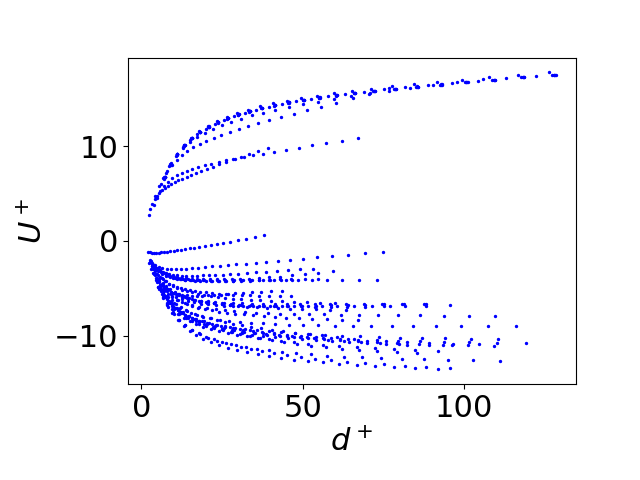}
\caption{Scatter plot of  $U^+$ vs. $d^+$.}
\label{target-scatter-uplus-yplus-diff}
\end{subfigure}
\caption{Diffuser flow. $d$ is the wall distance. \textcolor{blue}{$\bullet$}: location of cell centers. 
The dashed black line shows the contour of the lower wall (not to scale). Every second $x$ line and $d$ point are shown.}
\label{target-uplus-yplus-diff}
\end{figure}

A database for \kdtree\ is created by making a simulation of the flow in a diffuser, see Fig.~\ref{diffuser-domain}. 
The diffuser flow is selected because it is
a relatively simple test case including a flow region with adverse-pressure gradient. 
Compared to channel flow and flat-plate boundary layer flow, it adds
the complexity of a small re-circulation and $dP/dx >   0$. At first, a well-resolved
 LES was made with a grid of $600\times 150\times 300$ cells with  $0.3< \Delta y^+ <22 $ ($\Delta z^+ =11$, $\Delta x^+ =22$ at the inlet) 
at a low Reynolds number ($Re_\tau = 2\,000$
at the inlet). The disadvantage of this procedure is that we cannot require an IDDES with wall functions to perform as well as a well-resolved LES.
In the former case, inaccuracies may appear not only due to the wall functions but also due to insufficient resolution and errors
due to the turbulence model (i.e. IDDES).

Instead, we use WR-IDDES (i.e. wall-adjacent cells at $y^+ < 1$) 
with the underlying AKN $\ke$ turbulence model (see Sections~\ref{IDDES} and ~\ref{AKN}) when creating the database. The advantage 
is that when validating the \kdtree\ wall function, we can use an
identical grid in the wall-parallel planes. In this case it is reasonable
to require that the wall-functions should give similar results  as the  WR-IDDES.

Figure~\ref{diffuser-domain} shows the domain of the diffuser which is used for creating the first database.
All length are made non-dimensional using the inlet height, $h$.
 The opening angle is $15^o$. The maximal
height is $H_{max} = 3.4$.  The length of the straight channel downstream of the inlet is  $L_1=6$.
The extent of the diffuser and the following straight channel  is $L_2=17.5$.
The length of the weak contraction is $L_3=22.5$;  the contraction is used in order to avoid backflow at the outlet. 
The friction Reynolds number at the inlet is $Re_\tau = 5\, 200$ based on the friction velocity and the inlet height, $h$.

The boundary conditions are set as follows. The top and lower boundaries are  slip and no-slip wall, respectively.
At the wall, $\ub=\vb=\wb=k=0$ and the dissipation is given by Eq.~\ref{chienBC}.  Periodic boundary conditions are set
for all variables in the spanwise direction ($z$). Homogeneous Neumann boundary conditions are used for all variables at the outlet.
A pre-cursor WR-IDDES of channel flow (half-channel width)  is used for setting
the instantaneous inlet boundaries conditions for all variables except the pressure.
 The boundary condition for pressure at the four boundaries
in Fig.~\ref{diffuser-domain} is homogeneous Neumann. The grid has $700\times 90\times 96$ in $x$ (streamwise),
$y$ (wall-normal) and $z$ (spanwise) directions. 

Figure~\ref{cp-cf-vel-diff} shows the predicted pressure coefficient, skin friction and velocity profiles at 
$x=5$, $11$, $18$, $21$ and $x=24$.
The diffuser creates a pressure increase of approximately half a dynamic pressure height and the 
flow is separated between  $x=3.15$ and $x=27$.


The created database consists of $41$  time-averaged profiles of $U^+$ vs. $d^+$ with $26$ points in each profile, see 
Fig.~\ref{target-x-yplus-diff} ($d$ denotes the wall distance).
 The bullets show every second $d$ location. The vertical lines show the $x$ location
of the profiles and the extent of these lines show the largest $d^+$ at each $x$; the largest wall distance is approximately 
the same for each $x$ location since the largest $d$ is defined by a wall-parallel grid line. 
Figure~\ref{target-scatter-uplus-yplus-diff} shows that the 
magnitudes of the separated (negative) $U^+$ velocity profiles are similar to those of the attached profiles. The reason
for the large magnitude of the separated velocity profiles is that the friction velocity, $u_\tau$, is small in this region.

\begin{figure}
\centering
\begin{subfigure}[t]{0.5\textwidth}
\begin{tikzpicture}[xscale=0.8, yscale=0.8]
\begin{axis}
[x=1cm,y=1cm,xmin=0,xmax=7,ymin=0,ymax=1,
axis x line=none,
axis y line=none
]
\addplot [mark=none,fill=gray]file {contour-iddes-hump.txt};
\end{axis}
%
\draw [myarrow-myarrow] (0,-1.0) -- (2.1,-1.0);
\node [above] at (1.05,-1.0) {$L_1$};

\draw [myarrow-myarrow] (2.1,-1) -- (3.1,-1.0);
\node [above] at (2.6,-1.0) {$c$};

\draw [myarrow-myarrow] (3.1,-1.0) -- (6.20,-1.0);
\node [above] at (4.65,-1.0) {$L_2$};

\draw [myarrow-myarrow] (2.1,0.4) -- (2.1,0) -- (2.5,0);
\node [below right] at (2.3,0.1) {$x$};
\node at (2.0,0.5) {$y$};

\node [above] at (2,0.8) {slip wall};

\node [above] at (5,-0.1) {wall};

\node [left] at (0.0,0.5) {$H$};
\node [above] at (3.2,0.0) {$h$};
\end{tikzpicture}
\caption{Geometry. Height of hump, $h=0.128$. $H=0.909$,  $L_1 = 2.1$, $L_2=4.1$.}
\label{hump-domain}
\end{subfigure}%
\hfill
\begin{subfigure}[t]{0.5\textwidth}
\centering
\includegraphics[scale=0.3,clip=]{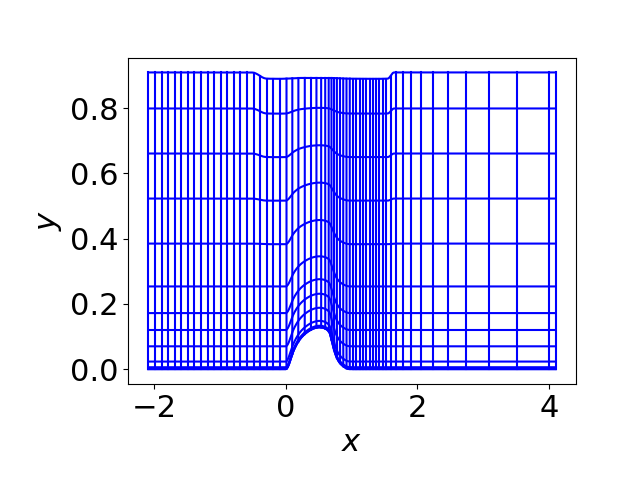}
\caption{Grid, $x-y$ plane. Every $10^{th}$ grid line is shown.}
\label{grid-hump}
\end{subfigure}
\caption{Hump flow.}
\end{figure}

\begin{figure}
\begin{center}
\begin{tikzpicture}[xscale=0.8, yscale=0.8]
\draw [thick] (0, 0) rectangle (6, 3);

\fill[color=lightgray,opacity=0.8] (0,0) rectangle (2,3);
\node [rotate=90] at (1,1.5) {$\ub k_{tot}  \frac{\D f_k}{\D x} < 0$};

\draw [myarrow-myarrow]  (0,3.4) -- (2,3.4);
\node [above] at (1,3.4) {$\Delta x$};

\draw [-myarrow] (-1.5, 1.5) .. controls(-1.3,2.1) and (-1,0.8)   .. (-0.5, 1.5);
\node [above right] at (-2.0,1.5) {$\ub^{\prime},\vb^{\prime},\wb^{\prime}$};

\draw [-myarrow] (-1.5, 2.5) .. controls(-1.3,3.1) and (-1,1.8)   .. (-0.5, 2.5);

\draw [-myarrow] (-1.5, 0.5) .. controls(-1.3,1.1) and (-1,-0.2)   .. (-0.5, 0.5);

\node [rotate = 90] at (0.2,1.5) {inlet};

\node at (3.0,0.2) {wall};

\node at (3.0,3.2) {wall};

\draw [myarrow-myarrow] (-1.5,0) -- (-1.5,-1) -- (-0.5,-1);
\node [below right] at (-0.8,-1) {$x$};
\node [left] at (-1.5,0) {$y$};

\end{tikzpicture}
\end{center}
\caption{Setting inlet boundary conditions. Synthetic inlet fluctuations~\citep{shur:14}, $\ub', \vb' \wb'$  are superposed
on a mean RANS inlet velocity profile. The commutation term, $\ub \D f_k/\D x$ (see Eq.~\ref{comm-fk}), is added to the first cell layer.}
\label{inlet-bc}
\end{figure}

\begin{figure}
\centering
\begin{subfigure}[t]{0.33\textwidth}
\centering
\includegraphics[scale=0.28,clip=]{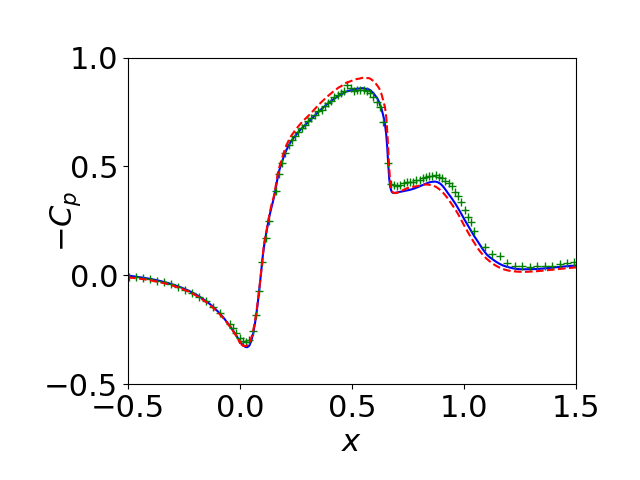}
\caption{Pressure coefficient.}
\end{subfigure}%
\hfill
\begin{subfigure}[t]{0.33\textwidth}
\centering
\includegraphics[scale=0.28,clip=]{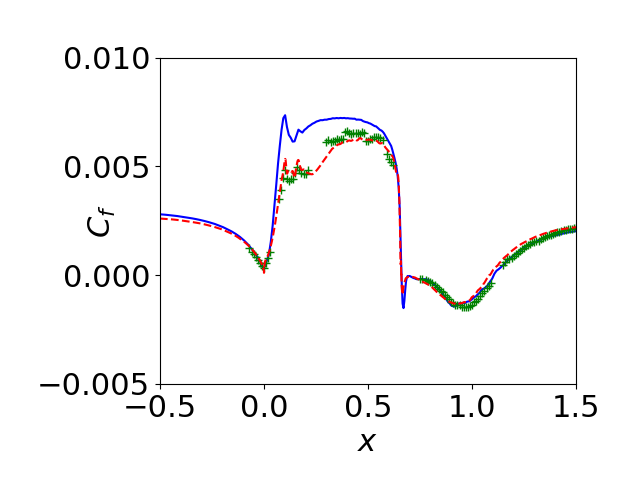}
\caption{Friction coefficient.}
\end{subfigure}%
\hfill
\begin{subfigure}[t]{0.33\textwidth}
\centering
\includegraphics[scale=0.28,clip=]{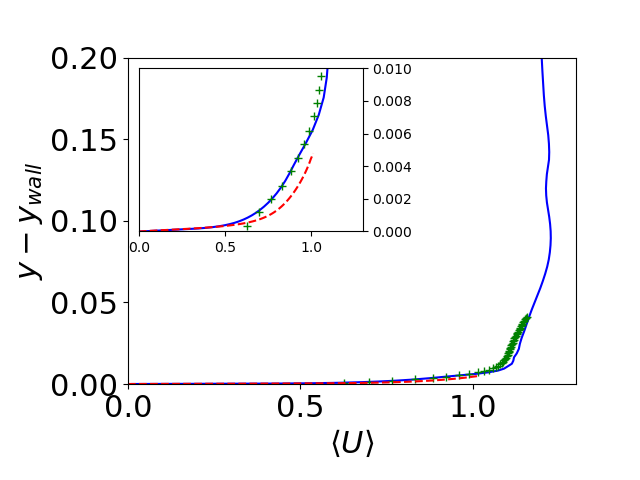}
\caption{Velocity at $x=0.65$.}
\end{subfigure}
\begin{subfigure}[t]{0.33\textwidth}
\centering
\includegraphics[scale=0.28,clip=]{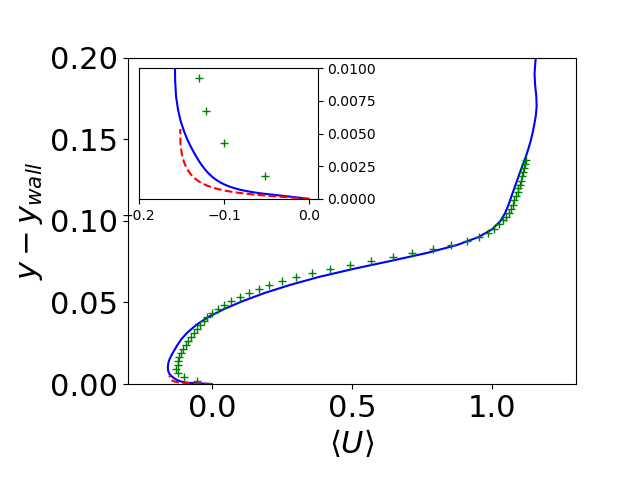}
\caption{Velocity at $x=0.80$.}
\end{subfigure}%
\begin{subfigure}[t]{0.33\textwidth}
\centering
\includegraphics[scale=0.28,clip=]{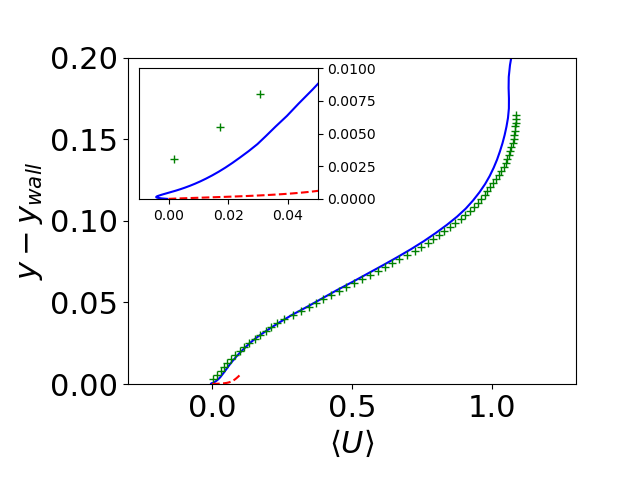}
\caption{Velocity at $x=1.10$.}
\end{subfigure}%
\begin{subfigure}[t]{0.33\textwidth}
\centering
\includegraphics[scale=0.28,clip=]{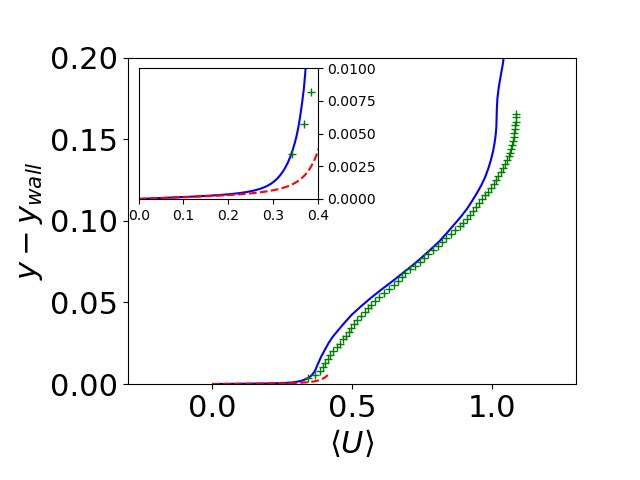}
\caption{Velocity at $x=1.30$.}
\end{subfigure}%
\caption{Hump flow. \solidline: WR-IDDES, see Fig.~\ref{lowre-grid};
\dashedline: WR-LES~\citep{uzun:18,uzun:www};
\textcolor{green}{$+$}: experiments~\citep{greenblatt:04}.}
\label{hump-low-IDDES}
\end{figure}

\begin{figure}
\centering
\begin{subfigure}[t]{0.33\textwidth}
\centering
\includegraphics[scale=0.28,clip=]{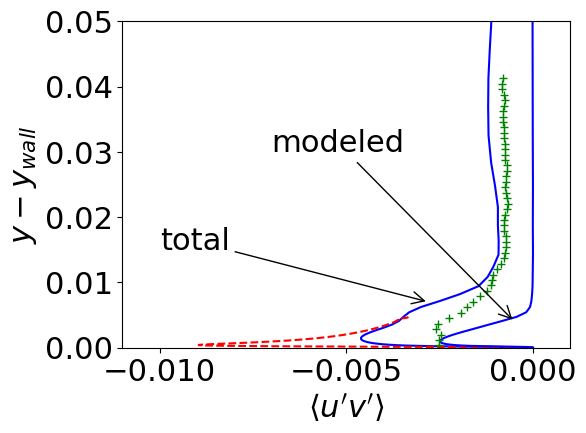}
\caption{$x=0.65$.}
\end{subfigure}%
\begin{subfigure}[t]{0.33\textwidth}
\centering
\includegraphics[scale=0.28,clip=]{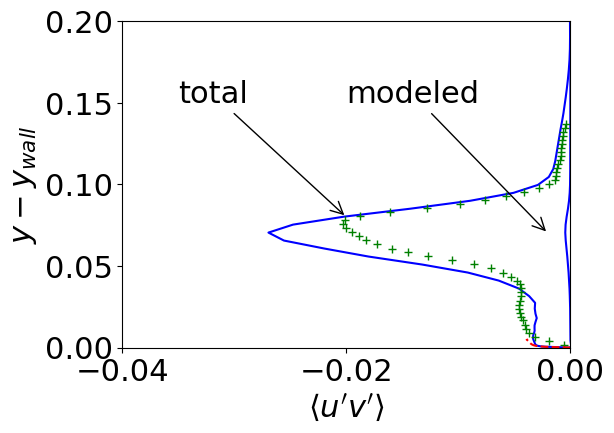}
\caption{$x=0.80$.}
\end{subfigure}
\begin{subfigure}[t]{0.33\textwidth}
\centering
\includegraphics[scale=0.28,clip=]{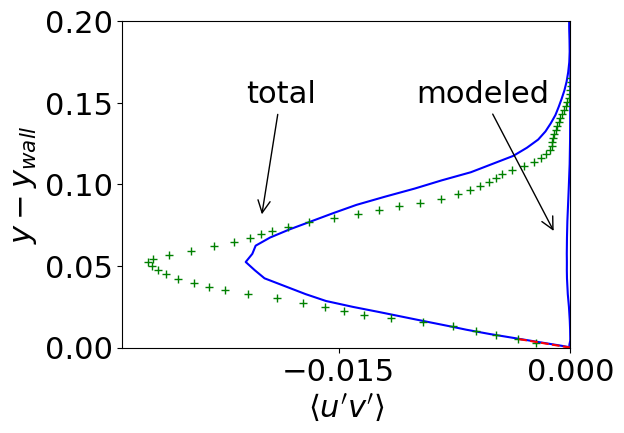}
\caption{$x=1.10$.}
\end{subfigure}%
\begin{subfigure}[t]{0.33\textwidth}
\centering
\includegraphics[scale=0.28,clip=]{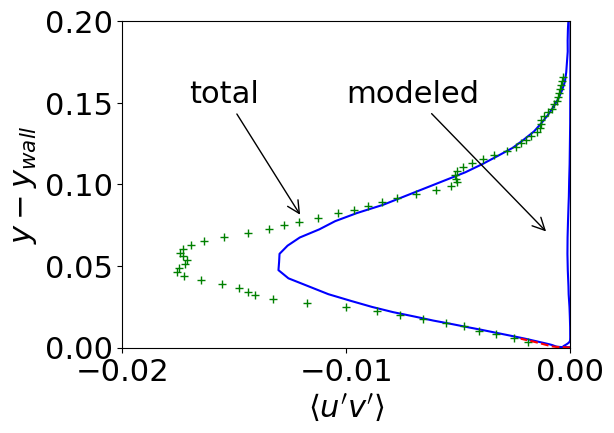}
\caption{$x=1.30$.}
\end{subfigure}%
\caption{Hump flow.  Total turbulent shear stress. \solidline: WR-IDDES, see Fig.~\ref{lowre-grid};
\dashedline: WR-LES~\citep{uzun:18,uzun:www};
\textcolor{green}{$+$}: experiments~\citep{greenblatt:04}.}
\label{hump-uv-lowre}
\end{figure}

\begin{figure}
\centering
\begin{subfigure}[t]{0.5\textwidth}
\centering
\includegraphics[scale=0.30,clip=]{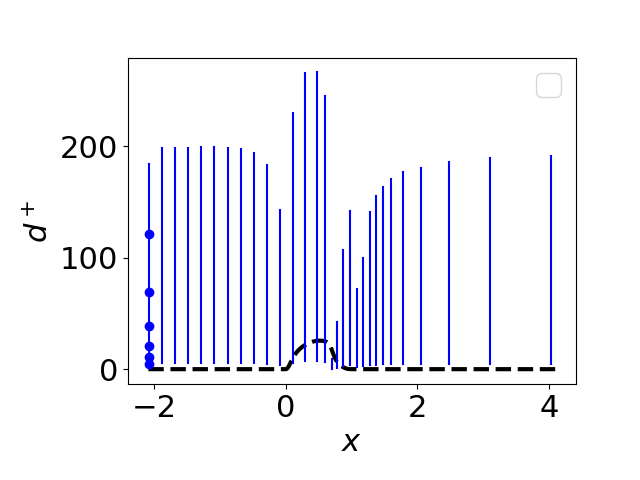}
\caption{Target data points of $d^+$ vs. $x$.}
\label{target-x-yplus-hump}
\end{subfigure}%
\hfill
\begin{subfigure}[t]{0.5\textwidth}
\centering
\includegraphics[scale=0.30,clip=]{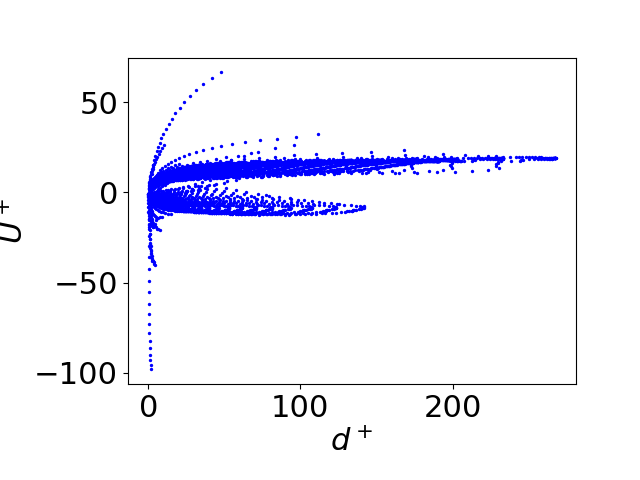}
\caption{Scatter plot of  $U^+$ and $d^+$.}
\label{target-scatter-uplus-yplus-hump}
\end{subfigure}
\caption{Hump flow. $d$ is the wall distance. The dashed black line shows the contour of the lower wall (not to scale).
 The target database consists of time-averaged $582$ profiles (all grid lines)
of $U^+$ vs. $d^+$ with $24$ points in each profile.
Every $20^{th}$  $x$ line and every $4^{th}$ $d$ point are shown.}
\label{target-uplus-yplus-hump}
\end{figure}

\begin{figure}
\centering
\begin{subfigure}[t]{0.33\textwidth}
\centering
\includegraphics[scale=0.28,clip=]{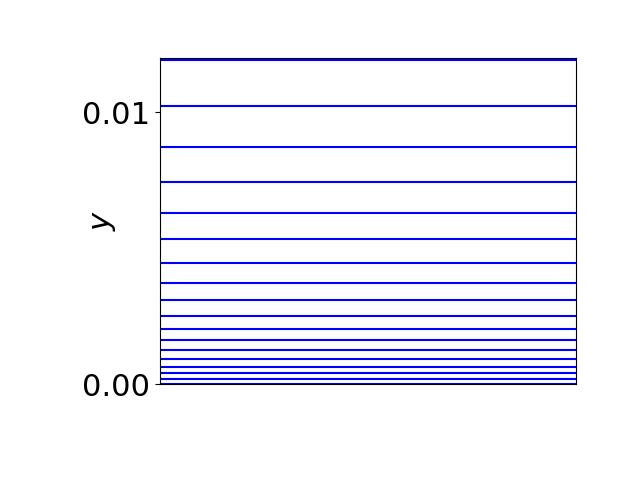}
\caption{WR-IDDES grid.}
\label{lowre-grid}
\end{subfigure}%
\hfill
\begin{subfigure}[t]{0.33\textwidth}
\centering
\includegraphics[scale=0.28,clip=]{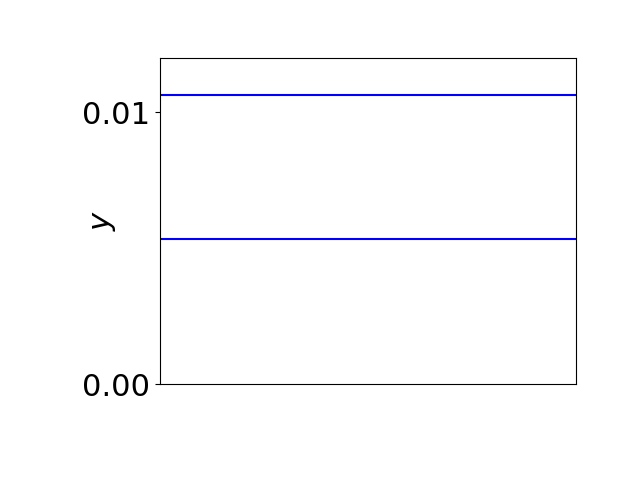}
\caption{Wall function grid.}
\label{wallf-grid}
\end{subfigure}%
\hfill
\begin{subfigure}[t]{0.33\textwidth}
\centering
\includegraphics[scale=0.28,clip=]{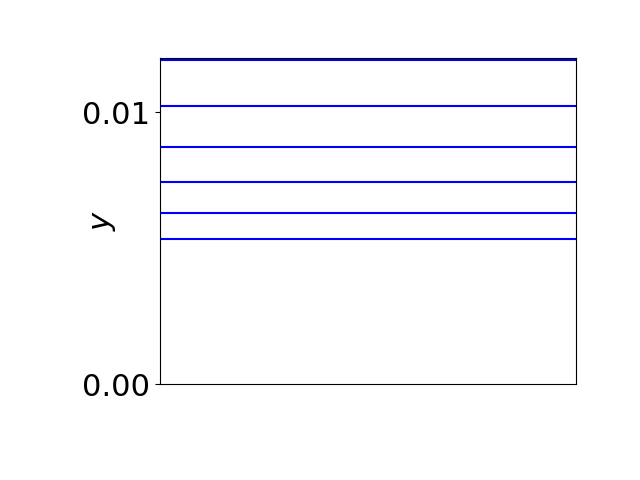}
\caption{New wall function grid.}
\label{new-wallf-grid}
\end{subfigure}
\caption{Different grids. \solidline: grid lines.}
\label{grid}
\end{figure}

A second target database is created by computing the flow over a hump.
The Reynolds number is $Re_c=936\, 000$ based on the inlet bulk velocity and the length of the hump, $c$. 
Figure~\ref{hump-domain} shows the geometry. All length and velocities are scaled with the hump length, $c$, 
and the inlet bulk velocity,
respectively.
It is a two-dimensional hump mounted between two glass end plate frames and both leading edge and trailing edges are 
faired smoothly with a wind tunnel splitter plate.
The experiments were carried out at NASA~\citep{greenblatt:04,naughton:06a} (see also \cite{nasa_workshop:04}). It is a challenging
flow involving developing boundary layer, accelerating flow, separation and a recovery region.

This test case has been used in both \cite{nasa:workshop} workshop as well as EU projects such as 
ATAAC~\citep{ATAACK-book} and Go4Hybrid~\citep{hump:go4hybrid:18}. \cite{uzun:18} made wall-resolved LES using an SGS model
by~\cite{vreman:04}.
The largest computational grid had $850$ million cells. They report good agreement with experiments.

The spanwise extent is $z_{max}=0.2$.
The grid has  $582 \times 128 \times 64$ cells ($x$, $y$, $z$), see Fig.~\ref{grid-hump}.
The mean inlet velocity is taken from a pre-cursor 
2D RANS of a flat-plate boundary layer. Instantaneous turbulent fluctuations are superimposed
on the mean RANS velocity profile. The inlet $k$ and $\e$ are also taken from the 2D RANS simulation.

In order to reduce the modeled turbulence to a suitable LES level a commutation 
term is added to the $k$ equation.
The method was proposed  for the PANS model~\citep{davidson:15a}. Although the PANS model is not used
in this work is was shown in \cite{friess-davidson:20} that IDDES and PANS can be formulated in a way so that
PANS and IDDES are equivalent because of the equivalence criterion~\citep{friess:15}.
In PANS, the equation for the modeled turbulent kinetic energy, $k$,
 is derived by multiplying the $k_{tot}$ equation ($k_{tot}= k_{res}+k$, where $k_{res}$ denotes resolved, turbulent kinetic energy) 
by $f_k$ where $f_k=k/k_{tot}$ 
($f_k$ denotes the ratio of modeled to total turbulent kinetic energy and is one in RANS regions and smaller than one 
($0.2 \lesssim f_k \lesssim 0.6$) in LES regions).
The convective term in the $k$ equation  -- i.e. the left-hand side of the equation -- with 
constant $f_k$ is then obtained as ($D/Dt$ denotes the material derivative)
\begin{eqnarray}
f_k \frac{Dk_{tot}}{Dt}= \frac{ D (f_kk_{tot})}{Dt} = \frac{Dk}{Dt}
\end{eqnarray}
Now, if $f_k$ varies in space, we get instead
\begin{eqnarray}
\label{kmodel-source}
f_k \frac{Dk_{tot}}{Dt}=
\frac{D (f_k k_{tot})}{Dt} -   k_{tot} \frac{Df_k}{Dt}
=  \frac{D k}{Dt} -   k_{tot} \frac{Df_k}{Dt}
\end{eqnarray}
Consider this  equation near the inlet. Since the inlets in the present are located at the  low $x$ end of the domain 
(see Fig.~\ref{inlet-bc})
and $f_k$ is constant in time, $Df_k/Dt = \ub \D f_k/\D x$. 
The last term in Eq.~\ref{kmodel-source} is the commutation term which is
 discretized as
\begin{eqnarray}
\label{comm-fk}
k_{tot} \ub \frac{\D f_k}{\D x} = k_{tot} \ub \frac{f_{k,0.5\Delta x} - f_{k,x=0}}{0.5\Delta x}
\end{eqnarray}
where $\Delta x$ denotes the distance between the inlet and the cell center adjacent to the inlet and the inlet is located
at $x=0$, see Fig.~\ref{inlet-bc}.
The value of $f_k$ at the inlet is one ($k_{RANS}$  is prescribed as inlet boundary condition) and $f_k$ at the cell center,
$x=0.5\Delta x$, is computed as \citep{friess-davidson:20}
\begin{eqnarray}
\label{fx-0.5dx}
f_k= \min \left[ 1,   \max \left( 0 , \left( \frac{ \ceii - \cei \psi }{C_{\e 2}-C_{\e 1}} \right)^{1/3} \right) \right]
\end{eqnarray}
where $\psi$ is given by Eq.~\ref{psi-eq}. $f_k$ in Eq.~\ref{fx-0.5dx} is smaller than one (because $x=0.5\Delta$ is 
in the LES region because of the synthetic, fluctuating  inlet boundary conditions)  and hence the commutation
term in Eq.~\ref{comm-fk} is positive on the left-hand of the $k$ equation and negative on the right-hand side thereby
reducing $k$ near the inlet from RANS values to LES values. The resolved part of $k_{tot}$ in Eq.~\ref{comm-fk} 
is computed as a running average.
  No commutation term is added to the $\e$ equation since the modeled dissipation, $\e$,  is the same in RANS and LES regions. 

The created database consists of time-averaged profiles of $U^+$ vs. $d^+$ at all $x$ grid lines with $26$ points in each profile, see 
Fig.~\ref{target-x-yplus-hump}. The minimum ($\simeq -100$) and maximum ($\simeq 70$) 
data points of $U^+$ in Fig.~\ref{target-scatter-uplus-yplus-hump} stem from the separation
region at $x\simeq 0.7$ where $u_\tau$ is very small. Many more database points are used for this flow than for the diffuser
flow (see Fig.~\ref{target-x-yplus-diff}) because this flow is much more complex.

The predicted pressure coefficient, 
skin friction and velocity profiles are presented at 
$x=0.65$, $0.80$, $1.10$ and $x=1.30$, see 
Fig~\ref{hump-low-IDDES}. The agreement with experiments is good. However,
the predicted total, turbulent shear stress is over-predicted by almost 
a factor of two; the agreement 
further downstream is better. The predicted shear stresses, see Fig.~\ref{hump-uv-lowre},  are similar to those reported
in~\cite{hump:go4hybrid:18}. The WR-LES data by~\cite{uzun:18,uzun:www} are also shown.
The WR-LES data using the fine grid, wide domain and modified top wall contour are used.  
Note that data only for $y-y_{wall} < 0.03$ are available. It can be seen that the skin friction for $0.1 < x < 0.3$ is much better
predicted with WR-LES than WR-IDDES. The flow is accelerating in this region and the low $C_f$ indicates that the turbulence is
perhaps re-laminarizing. \cite{uzun:18} report that this indeed is the case since the  relaminarization parameter, 
$K \simeq 4.87\cdot 10^{-6}$,
is larger than the value $K= 3\cdot 10^{-6}$ at which relaminarization takes place. Although the WR-LES predicts
the skin friction and the pressure coefficient in very good agreement with experiments
 it over-predicts the turbulence shear stress at $x=0.65$ by a factor of three.

In \cite{davidson:23-url} a new grid strategy was proposed for wall-functions grids
(it was also used in~\cite{paik:12}). Figure~\ref{grid} shows three grids.
A WR-IDDES grid (Fig.~\ref{lowre-grid}), a standard wall-function grid (Fig.~\ref{wallf-grid}) and the new proposed 
wall-function grid (Fig.~\ref{new-wallf-grid}). The center of the first cell in a WR-IDDES grid is located at $y^+ < 1$
and then the wall-normal cell size increases by approximately $10\%$ away from the wall.
In a standard wall-function grid, the center of the first cell is located in the logarithmic region, say,  $30 < y^+ < 400$, and
then the cells may increase by $10\%$ away from the wall (or the cell size may stay constant). In the new wall-function
grid strategy, a number of the cells 
in the WR-IDDES grid (Fig.~\ref{lowre-grid}) are merged into one large cell. Further away from the wall, the grid is identical to 
the WR-IDDES grid. The advantage of this new grid is that the velocity gradients away from wall (i.e. at $y > 0.005$,
see Fig.~\ref{new-wallf-grid}) are resolved as accurately as at the WR-IDDES grid. Wall function grids 
such as that in Fig.~\ref{new-wallf-grid} are used unless otherwise stated.

\section{Machine Learning}

\label{ML:sec}

We have two sets of data points. One is the target data set ($\mathbf{X} =[U^+_{target}, y^+_{target}]$) and the other one is the
CFD data set ($\mathbf{x}=[U^+_{CFD}, y^+_{CFD}]$). The target data set is either taken from the diffuser flow 
(see Fig.~\ref{target-uplus-yplus-diff}) or from the hump flow (see Fig.~\ref{target-uplus-yplus-hump}).
The CFD data set  includes all wall-adjacent cells in the CFD simulation. The Python
\kdtree\ is used which computes the distance between the vectors as 
\begin{equation}
\label{d_s}
\mathbf{d_s} = \mathbf{X}_i - \mathbf{x}_j  
\end{equation}
(see Line 49 in Listing~\ref{ML-code} in Appendix~\ref{appendix-kdtree})
for all $j$ and finds the $K$ smallest distances for each $j$. 
Indices $i$ and $j$ denote the row of $\mathbf{X}$ and $\mathbf{x}$, respectively, which both have two columns.
$\mathbf{X}_i$ (the database) is set at lines 21-24 in Listing~\ref{ML-code} and
$\mathbf{x}_j$ (the CFD values) is set at lines 44, 45 in Listing~\ref{ML-code}.
In the present work we use $K=1$  or
$K=5$. For $K=1$, only the closest cell is used and for $K=5$ the target 
is computed by averaging over the five closest points using the weight $1/|\mathbf{d_s}|$ (see Eq.~\ref{d_s}). 
Unless otherwise stated, $K=1$ is used 
 for \kdtree\ with hump-flow data and $K=5$ is used for \kdtree\ with diffuser flow data. 
The impact of using $K=1$ or $K=5$ is usually small; when it does have an 
impact we will present predictions using both options. For the boundary-layer flow we will show the influence of using 
$K=1$, $2$, $5$ and $K=10$.

The Python module \texttt{fix\_k} in \pycalc\, in which the ML wall function is implemented,
is schematically given in Listing~\ref{ML-code}. 
It is called every CFD iteration at every time step.

The input parameters, $y^+$ and $U^+$, are taken either at the wall-adjacent cells (j\_wall =0, see Line 31 in Listing~\ref{ML-code}) 
or further away from the wall  (third cell) as suggested by \cite{larsson:12,larsson:16}. Unless otherwise stated,
the wall-adjacent cells are used.

As can be at the end of Listing~\ref{ML-code}, the output of this script is the friction velocity, $u_\tau$. 
It is used for setting the boundary conditions for the wall-parallel velocity, $k$ and $\e$ as
\begin{eqnarray}
\label{bc}
\rho u_\tau^2 &:& \ub \, \rm{equation}\nonumber\\
C_\mu^{-1/2} u_\tau^2 &:& k \, \rm{equation}\\
\frac{u_\tau^3}{\kappa y} &:& \e\,  \rm{equation}\nonumber
\end{eqnarray}
The boundary condition for the velocity is a shear stress boundary condition. The modeled turbulent kinetic energy, $k$,
 and its dissipation, $\e$,  are set at the wall-adjacent cells centers
according to Eq.~\ref{bc}.
The complete Python script in Listing~\ref{ML-code} and the databases can be found in \cite{ML-paper-code:24}.

It should be mentioned that databases with 
instantaneous data were also investigated (but no results are presented).
In that case $200$ data sets (independent in time) were created storing $U^+$ and $y^+$ at the same locations
as in Figs.~\ref{target-x-yplus-diff} and \ref{target-x-yplus-hump}. 
The instantaneous databases were hence $200$  times larger than the time-averaged ones.
We had hoped that instantaneous databases should be better than time-averaged data 
in  capturing large, unsteady characteristics of separated flow regions.
However, it was found that the instantaneous databases gave slightly worse results than the time-averaged ones.

\section{Standard wall functions}

\label{wall-functions}
The \kdtree\  wall functions will be compared to wall functions based on 
\reich
\begin{eqnarray}
\label{reich}
\frac{\ub_P}{u_\tau} \equiv U_P^+ =  \nonumber  \\
\frac{1}{\kappa} \ln(1-0.4y_P^+)+7.8\left[ 1-\exp\left(-y_P^+/11\right) - 
(y_P^+/11)\exp\left(-y_P^+/3\right)\right]
\end{eqnarray}
The friction velocity is then obtained by re-arranging Eq.~\ref{reich} and solving the implicit equation
\begin{eqnarray}
\label{ustar-reich}
u_\tau  - \ub_P\left\{
\ln(1-0.4y_P^+)/\kappa+\right . \nonumber  \\
\left . 7.8\left[ 1-\exp\left(-y_P^+/11\right) - 
(y_P^+/11)\exp\left(-y_P^+/3\right)\right]\right\}^{-1} = 0
\end{eqnarray}
using the Newton-Raphson method \verb+scipy.optimize.newton+ in Python. 
$\ub_P$ and $y_P^+$ denote the  
wall-parallel velocity and non-dimensional wall distance, respectively,
at the first  or third
wall-adjacent cells. Unless otherwise stated, the first wall-adjacent cells are used.

\begin{figure}
\centering
\begin{subfigure}[t]{0.33\textwidth}
\centering
\includegraphics[scale=0.28,clip=]{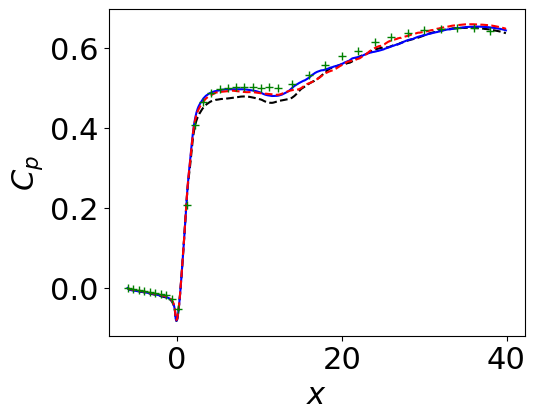}
\caption{Pressure coefficient.}
\end{subfigure}%
\hfill
\begin{subfigure}[t]{0.33\textwidth}
\centering
\includegraphics[scale=0.28,clip=]{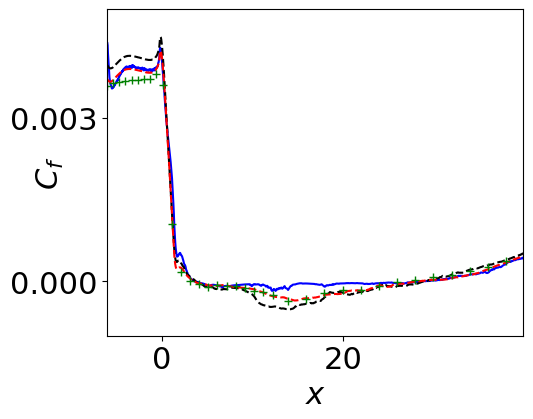}
\caption{Skin coefficient.}
\label{cf-u-diff-15-ni700}
\end{subfigure}%
\hfill
\begin{subfigure}[t]{0.33\textwidth}
\centering
\includegraphics[scale=0.28,clip=]{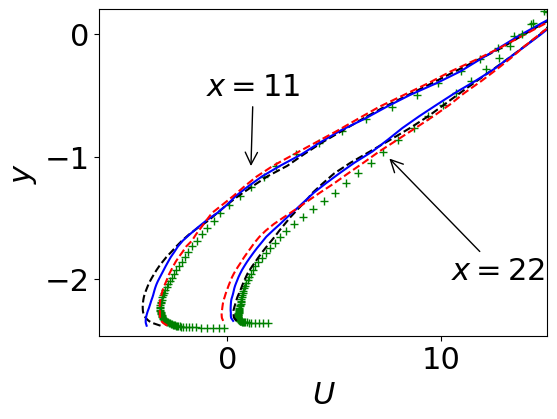}
\caption{Velocity profiles.}
\end{subfigure}
\caption{Diffuser flow, $\a=15^o$. $(N_x\times N_z) = (700\times 96)$. 
\solidline: \kdtree\ using hump flow data;
\dashedblackline: \kdtree\ using diffuser flow data;
  \dashedline: \wreich; \textcolor{green}{$+$}: WR-IDDES $(700\times 90 \times 96)$, see Fig.~\ref{lowre-grid}.}
\label{cp-cf-u-diff-15-ni700}
\end{figure}

\begin{figure}
\centering
\begin{subfigure}[t]{0.33\textwidth}
\centering
\includegraphics[scale=0.28,clip=]{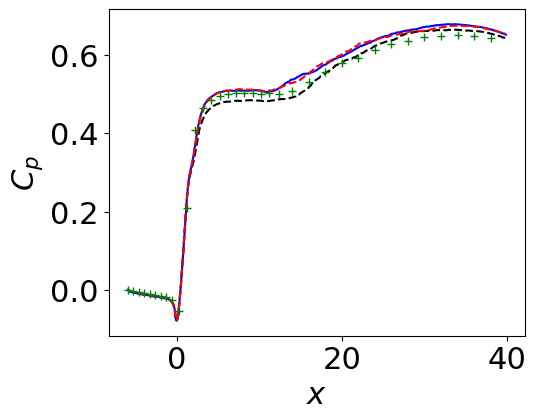}
\caption{Pressure coefficient.}
\end{subfigure}%
\hfill
\begin{subfigure}[t]{0.33\textwidth}
\centering
\includegraphics[scale=0.28,clip=]{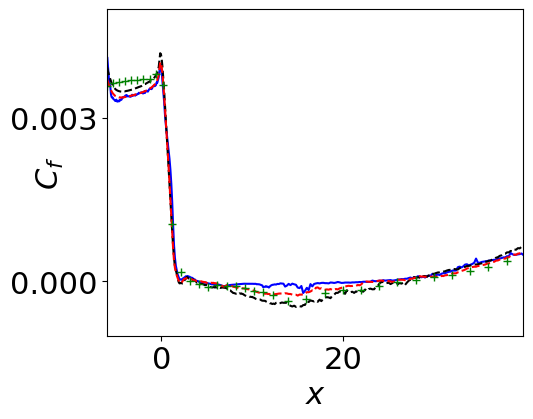}
\caption{Skin coefficient.}
\label{cf-u-diff-15}
\end{subfigure}%
\hfill
\begin{subfigure}[t]{0.33\textwidth}
\centering
\includegraphics[scale=0.28,clip=]{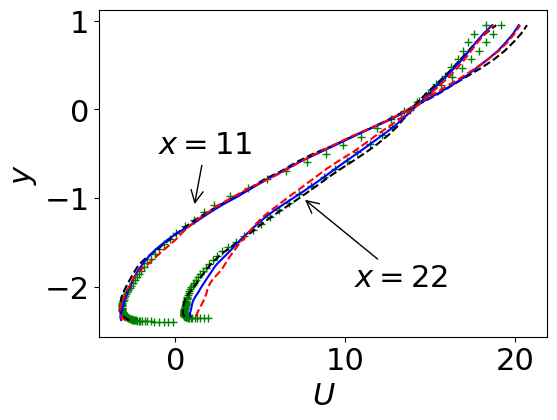}
\caption{Velocity profiles.}
\end{subfigure}
\caption{Diffuser flow, $\a=15^o$. $(N_x\times N_z) = (350\times 48)$. 
\solidline: \kdtree\ using hump flow data;
\dashedblackline: \kdtree\ using diffuser flow data;
  \dashedline: \wreich; \textcolor{green}{$+$}: WR-IDDES $(700\times 90 \times 96)$, see Fig.~\ref{lowre-grid}.}
\label{cp-cf-u-diff-15}
\end{figure}

\begin{figure}
\centering
\begin{subfigure}[t]{0.33\textwidth}
\centering
\includegraphics[scale=0.28,clip=]{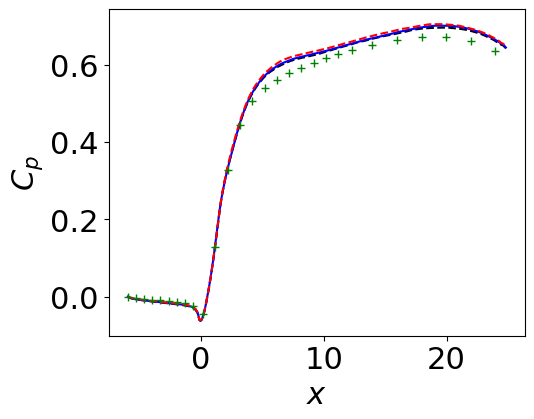}
\caption{Pressure coefficient.}
\end{subfigure}%
\hfill
\begin{subfigure}[t]{0.33\textwidth}
\centering
\includegraphics[scale=0.28,clip=]{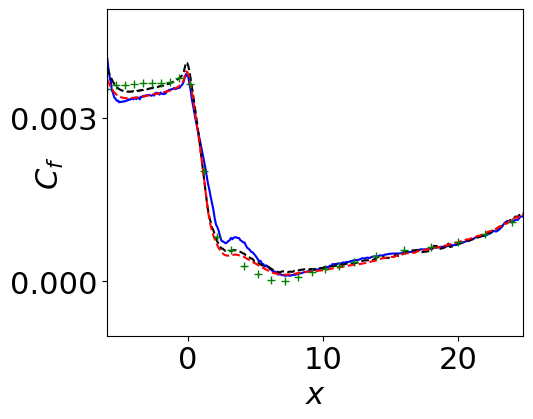}
\caption{Skin coefficient.}
\end{subfigure}%
\hfill
\begin{subfigure}[t]{0.33\textwidth}
\centering
\includegraphics[scale=0.28,clip=]{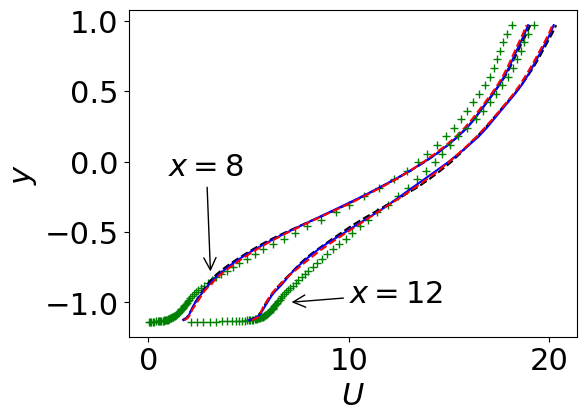}
\caption{Velocity profiles.}
\end{subfigure}
\caption{Diffuser flow, $\a=10^o$. $(N_x\times N_z) = (275\times 48)$. 
\solidline: \kdtree\ using hump flow data;
\dashedblackline: \kdtree\ using diffuser flow data;
  \dashedline: \wreich; \textcolor{green}{$+$}: WR-IDDES $(550\times 96\times 90)$, see Fig.~\ref{lowre-grid}.}
\label{cp-cf-u-diff-10}
\end{figure}

\section{Results}

\subsection{Diffuser flow}

Now we will validate the \kdtree\ wall functions. The diffuser flow with $15^o$ opening, i.e.
the same test case as the target data, is the first test case. 
The first $21$ cells near the wall are merged into one large
cell, see Fig.~\ref{new-wallf-grid}. We use the same grid in the
wall-parallel planes as the target data simulations (see Section~\ref{database}), i.e. 
$(N_x\times N_y \times N_z) = (700\times 70\times 96)$. 
At the inlet $y^+=35$ for the wall-adjacent cell center. The instantaneous inlet boundary
conditions ($\ub$, $\vb$, $\wb$, $k$ and $\e$) are taken from a precursor flow of fully-developed channel flow at $Re_\tau = 5\,200$
using IDDES with  wall functions based on \kdtree\ using hump flow data.

Next, the grid is coarsened by a factor two in both $x$ and $z$ direction compared
to the target-data simulations.
The domain is the same as in Fig.~\ref{diffuser-domain}. This gives a grid with
$350 \times 70\times 48$ cells. 
Finally, the diffuser angle is reduced to $\a = 10^o$ and a coarsened grid is used.  The contraction region is shorter (less chance of 
inflow at the outlet), $L_3 = 7$, which gives a grid with $275 \times 70\times 48$ cells.

Three wall functions are used: \kdtree\ using hump flow data (Section~\ref{ML:sec}), 
\kdtree\ using diffuser flow data (Section~\ref{ML:sec}) and \wreich\ (Section~\ref{wall-functions}). 
They are compared with WR-IDDES.
The predicted pressure coefficient, skin friction and two velocity profiles are shown in Figs.~\ref{cp-cf-u-diff-15-ni700}, 
\ref{cp-cf-u-diff-15} and \ref{cp-cf-u-diff-10}. Overall, the \kdtree\ using the diffuser flow data gives slightly better results
than the other two wall functions
and the \kdtree\ using the  diffuser flow data predicts somewhat too large a skin friction in the inlet region for the
fine mesh for $\a = 15^o$ and for $\a = 10^o$.
It may be noted that the skin frictions on the fine mesh (Fig.~\ref{cf-u-diff-15-ni700})
are larger than on the course mesh (Fig.~\ref{cf-u-diff-15}). The reason is that on the fine mesh, larger,  resolved turbulent fluctuations
are created by the synthetic inlet fluctuation because the finer mesh is able to resolve more turbulence than the coarse mesh.
For example, at $x=-5$ the magnitude of the peak absolute shear stress (not shown) is $1.15$ 
and $0.80$  at the fine and coarse mesh, respectively,
for \kdtree\ using diffuser flow data.

\begin{figure}
\centering
\begin{subfigure}[t]{0.33\textwidth}
\centering
\includegraphics[scale=0.28,clip=]{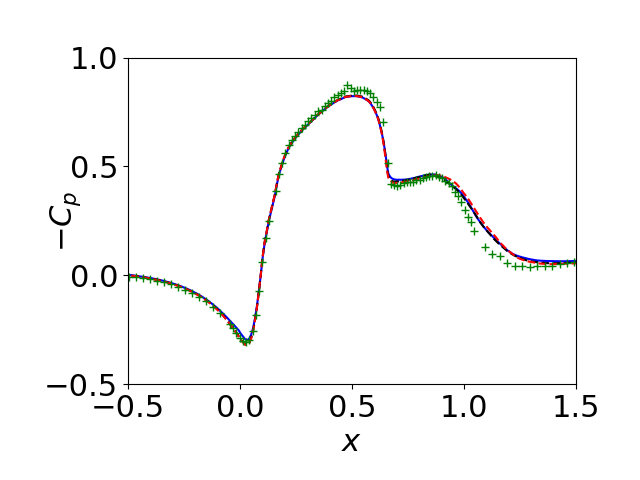}
\caption{Pressure coefficient.}
\end{subfigure}%
\hfill
\begin{subfigure}[t]{0.33\textwidth}
\centering
\includegraphics[scale=0.28,clip=]{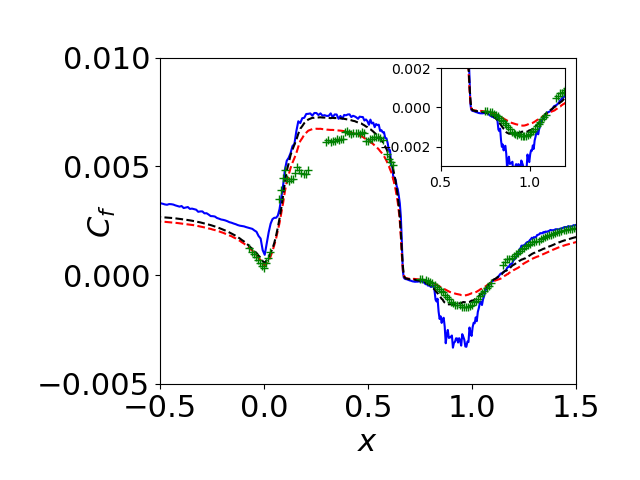}
\caption{Friction coefficient.}
\end{subfigure}%
\hfill
\begin{subfigure}[t]{0.33\textwidth}
\centering
\includegraphics[scale=0.28,clip=]{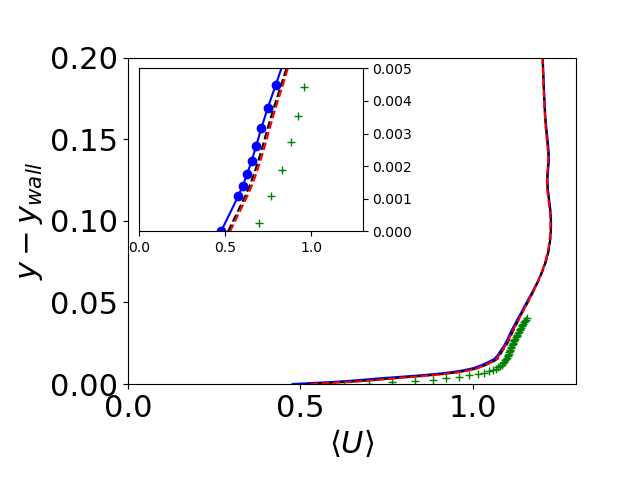}
\caption{$x=0.65$.}
\end{subfigure}
\begin{subfigure}[t]{0.33\textwidth}
\centering
\includegraphics[scale=0.28,clip=]{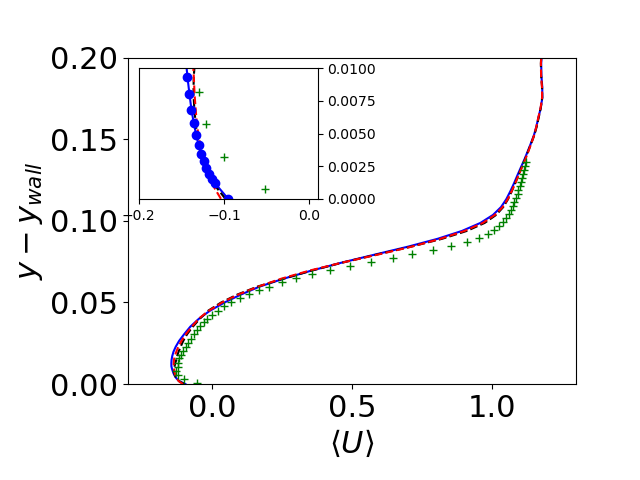}
\caption{$x=0.80$.}
\end{subfigure}%
\begin{subfigure}[t]{0.33\textwidth}
\centering
\includegraphics[scale=0.28,clip=]{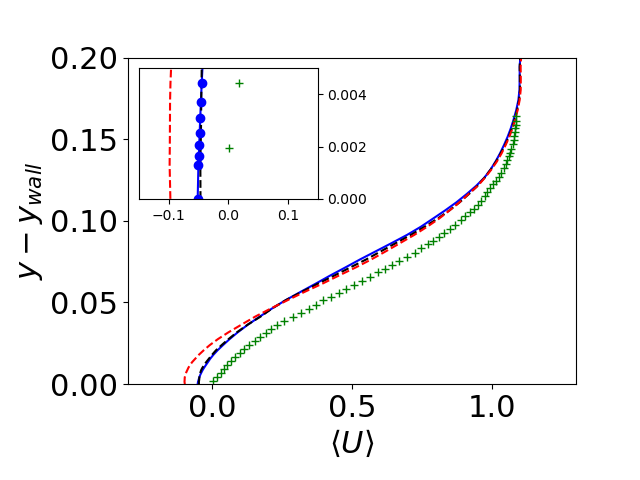}
\caption{$x=1.10$.}
\end{subfigure}%
\begin{subfigure}[t]{0.33\textwidth}
\centering
\includegraphics[scale=0.28,clip=]{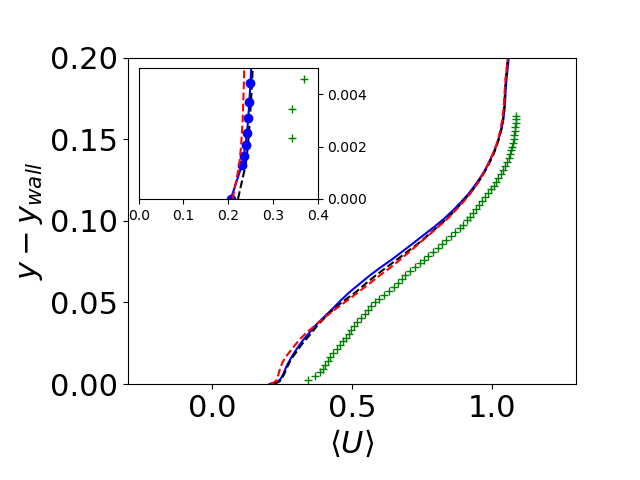}
\caption{$x=1.30$.}
\end{subfigure}%
\caption{Hump flow. $(N_x \times N_z) = (582\times 64)$. \solidline: \kdtree, hump flow data with $K=5$;  \dashedblackline: \kdtree, diffuser flow data;
  \dashedline: \wreich; \textcolor{blue}{$\bullet$}: location of cell centers;
\textcolor{green}{$+$}: experiments~\citep{greenblatt:04}.}
\label{hump-flow-ni583-nk-64}
\end{figure}

\begin{figure}
\centering
\begin{subfigure}[t]{0.33\textwidth}
\centering
\includegraphics[scale=0.28,clip=]{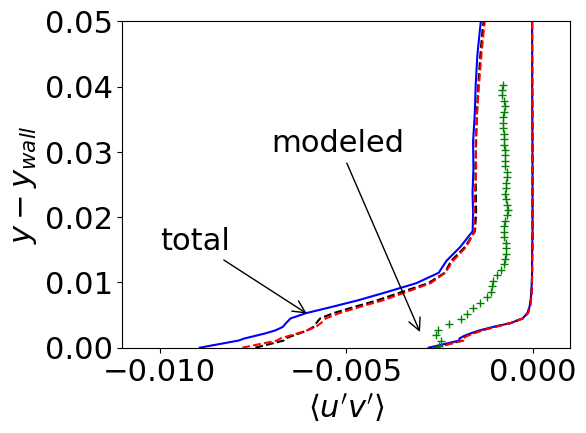}
\caption{$x=0.65$.}
\label{uv-hump-583}
\end{subfigure}%
\begin{subfigure}[t]{0.33\textwidth}
\centering
\includegraphics[scale=0.28,clip=]{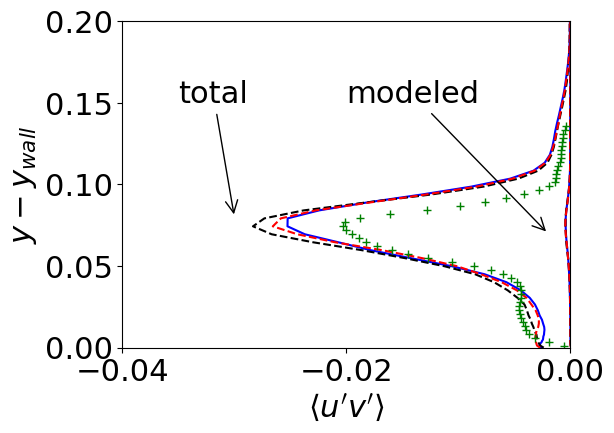}
\caption{$x=0.80$.}
\end{subfigure}
\begin{subfigure}[t]{0.33\textwidth}
\centering
\includegraphics[scale=0.28,clip=]{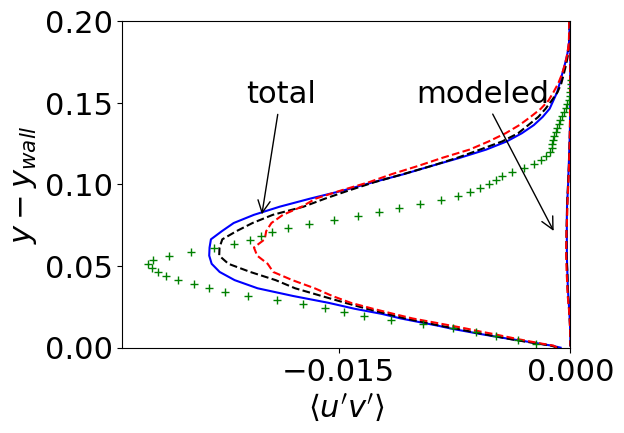}
\caption{$x=1.10$.}
\end{subfigure}%
\begin{subfigure}[t]{0.33\textwidth}
\centering
\includegraphics[scale=0.28,clip=]{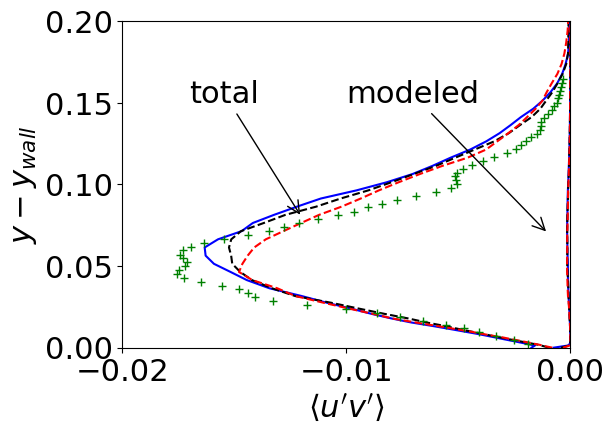}
\caption{$x=1.30$.}
\end{subfigure}%
\caption{Hump flow. $(N_x \times N_z) = (582\times 64)$.  Total turbulent shear stress. \solidline: \kdtree, hump flow data 
with $K=5$;  \dashedblackline: \kdtree, diffuser flow data;
  \dashedline: \wreich; 
\textcolor{green}{$+$}: experiments~\citep{greenblatt:04}.}
\label{hump-uv-ni583-nk-64}
\end{figure}

\begin{figure}
\centering
\begin{subfigure}[t]{0.33\textwidth}
\centering
\includegraphics[scale=0.28,clip=]{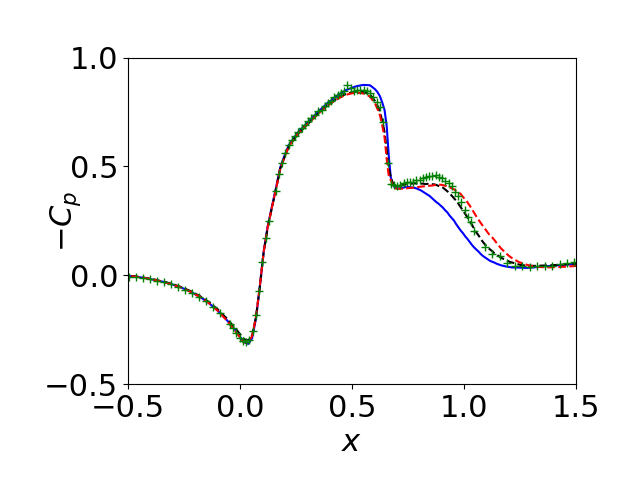}
\caption{Pressure coefficient.}
\end{subfigure}%
\hfill
\begin{subfigure}[t]{0.33\textwidth}
\centering
\includegraphics[scale=0.28,clip=]{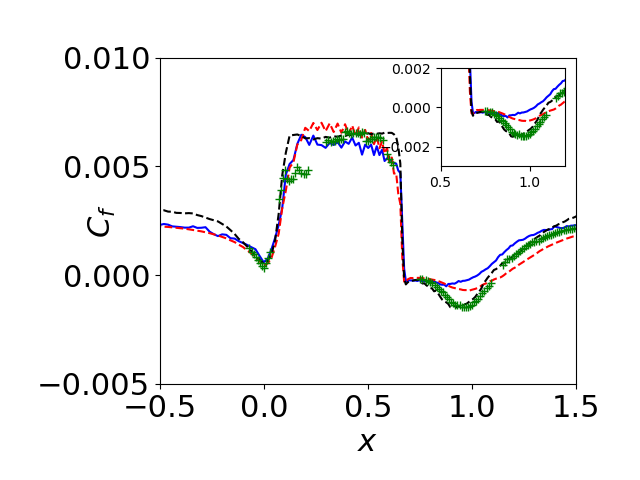}
\caption{Friction coefficient.}
\end{subfigure}%
\hfill
\begin{subfigure}[t]{0.33\textwidth}
\centering
\includegraphics[scale=0.28,clip=]{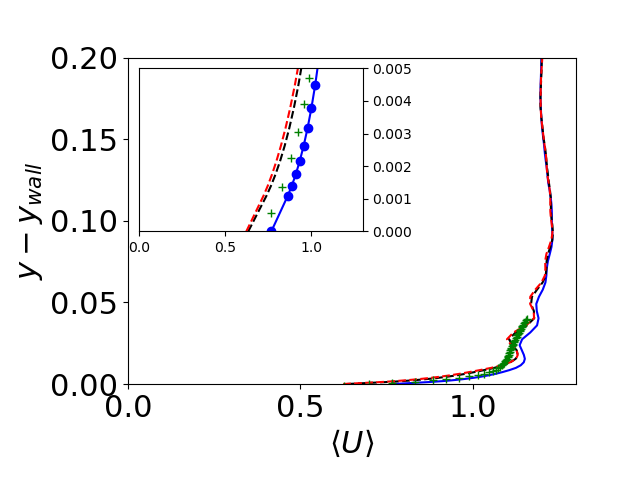}
\caption{$x=0.65$.}
\end{subfigure}
\begin{subfigure}[t]{0.33\textwidth}
\centering
\includegraphics[scale=0.28,clip=]{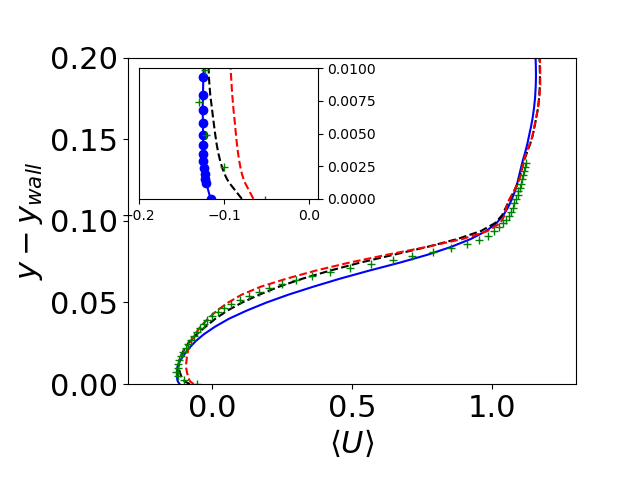}
\caption{$x=0.80$.}
\end{subfigure}%
\begin{subfigure}[t]{0.33\textwidth}
\centering
\includegraphics[scale=0.28,clip=]{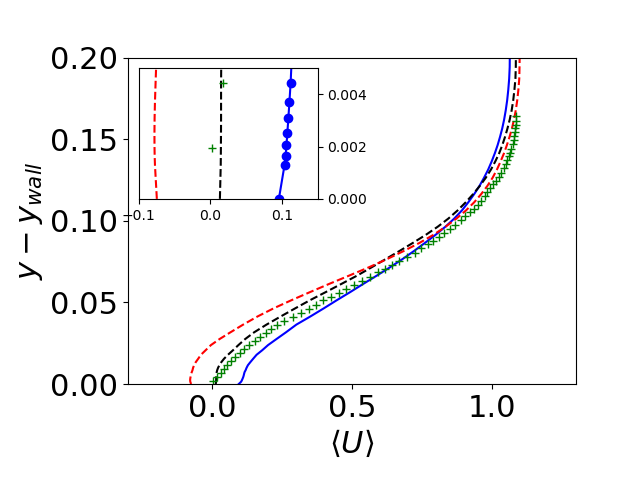}
\caption{$x=1.10$.}
\end{subfigure}%
\begin{subfigure}[t]{0.33\textwidth}
\centering
\includegraphics[scale=0.28,clip=]{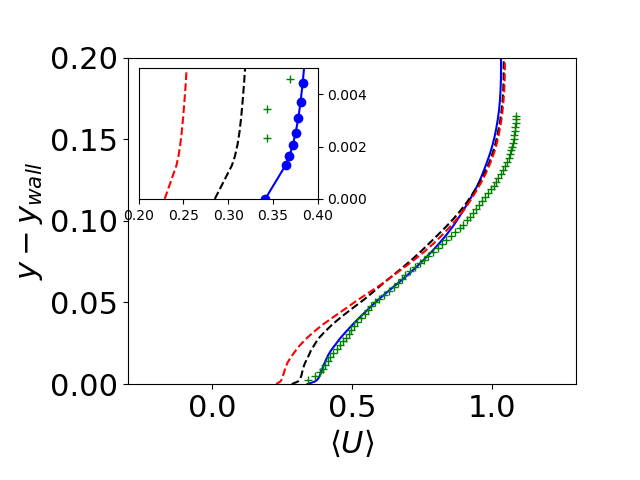}
\caption{$x=1.30$.}
\end{subfigure}%
\caption{Hump flow. $(N_x \times N_z) = (291\times 32)$. \solidline: \kdtree, hump flow data;  \dashedblackline: \kdtree, diffuser flow data;
  \dashedline: \wreich; \textcolor{blue}{$\bullet$}: location of cell centers;
\textcolor{green}{$+$}: experiments~\citep{greenblatt:04}.}
\label{hump-flow-ni291}
\end{figure}

\begin{figure}
\centering
\begin{subfigure}[t]{0.33\textwidth}
\centering
\includegraphics[scale=0.28,clip=]{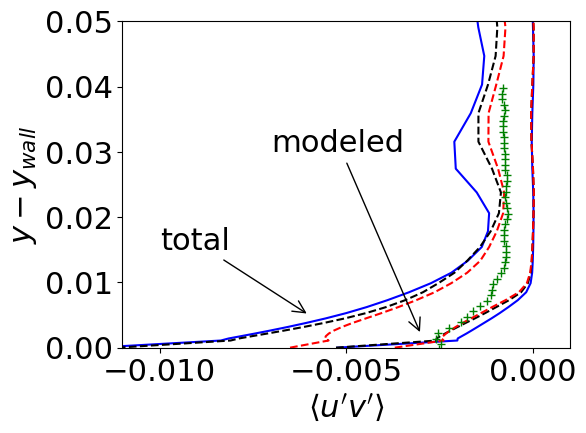}
\caption{$x=0.65$.}
\label{uv-hump-291}
\end{subfigure}%
\begin{subfigure}[t]{0.33\textwidth}
\centering
\includegraphics[scale=0.28,clip=]{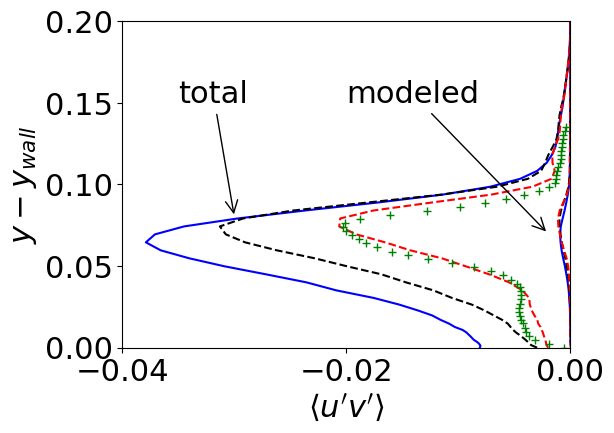}
\caption{$x=0.80$.}
\end{subfigure}
\begin{subfigure}[t]{0.33\textwidth}
\centering
\includegraphics[scale=0.28,clip=]{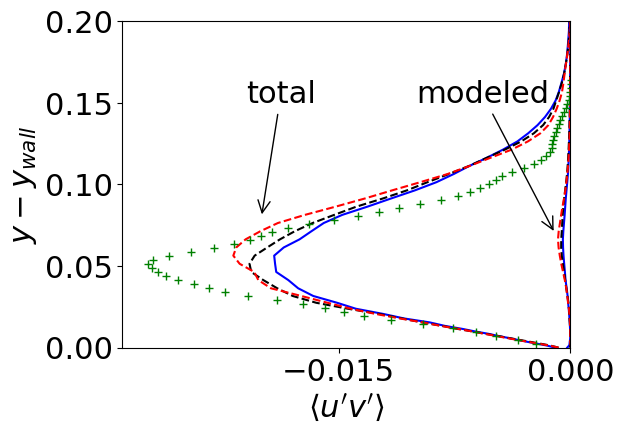}
\caption{$x=1.10$.}
\end{subfigure}%
\begin{subfigure}[t]{0.33\textwidth}
\centering
\includegraphics[scale=0.28,clip=]{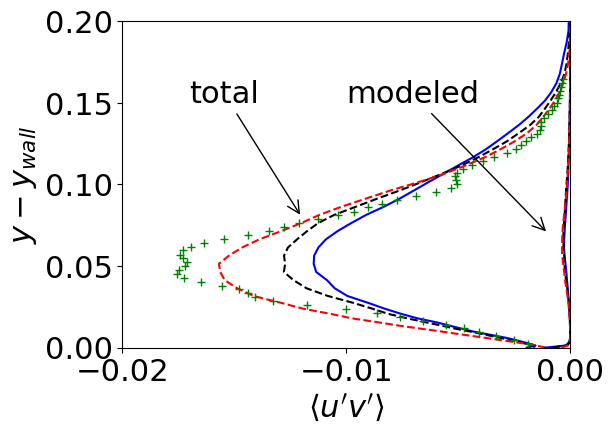}
\caption{$x=1.30$.}
\end{subfigure}%
\caption{Hump flow. $(N_x \times N_z) = (291\times 32)$.  Total turbulent shear stress. \solidline: \kdtree, hump flow data;  \dashedblackline: \kdtree, diffuser flow data;
  \dashedline: \wreich;
\textcolor{green}{$+$}: experiments~\citep{greenblatt:04}.}
\label{hump-uv-ni291}
\end{figure}

\subsection{Hump flow}

The inlet and outlet boundary conditions are the same as for the WR-IDDES simulations of the hump flow
 in Section~\ref{database} and Fig.~\ref{inlet-bc}.
Periodic boundary conditions are set in the spanwise direction.
First, we use the same grid in the $x-y$ plane as in the WR-IDDES.
The first $24$ cells near the wall are merged into one large cell. This  gives a grid with
$582\times 106 \times 64$ cells. 

The predictions using the three wall functions are presented in Figs.~\ref{hump-flow-ni583-nk-64}
and \ref{hump-uv-ni583-nk-64}.
Overall, the \kdtree\ with diffuser flow data shows the best agreement with 
experiments.  
The pressure suction peak for all wall functions is $4\%$ too low compared to experiments and WR-IDDES and
the velocity in the attached boundary layer at $x=0.65$  is approximately $2\%$ too low.
This is probably connected to the under-predicted velocity profiles at $x=1.10$ and $x=1.30$.
The skin friction on the upstream part of the hump is over-predicted by all wall functions. As mentioned when discussing
{Fig~\ref{hump-low-IDDES} above,
the turbulence in the experiment experiences a tendency to re-laminarization which all wall functions fail to capture.
The \kdtree\ with hump flow  data exhibits too strong a backflow when looking at the 
skin friction whereas the \wreich\ gives too weak a backflow. However, when looking at the velocity
profiles and shear stresses, all three wall functions 
give similar results. Furthermore, it is seen that the predicted 
shear stresses in the attached boundary layer ($x=0.65$) are much too large for all three wall functions. 
When compared with the stressed predicted with WR-IDDES
in Fig.~\ref{hump-uv-lowre} it is seen that it is the resolved stresses that are too large 
(the modeled one are of the same magnitude
in Figs.~\ref{hump-uv-ni583-nk-64} and \ref{hump-uv-lowre}). It should be noted that the peak of $\langle |u'v'|\rangle$ at $x=0.65$
in the WR-LES simulations
(Fig.~\ref{hump-uv-lowre}) is even larger than that in  Fig.~\ref{hump-uv-ni583-nk-64}.

In the next case, the grid is coarsened in both $x$ and $z$ directions, i.e. 
$291\times 106 \times 32$. 
Figures~\ref{hump-flow-ni291}
and \ref{hump-uv-ni291} present the results and the \kdtree\ with diffuser flow data actually gives
even better agreement than in Figs.~\ref{hump-flow-ni583-nk-64}
and \ref{hump-uv-ni583-nk-64}. This is a \enquote{lucky} coincidence due to a combination of modeled Reynolds stresses
and discretization errors.  Here $K=1$ for the \kdtree\ with hump flow data and it can 
be seen that the predicted skin friction is rather poorly predicted (very similar to that 
predicted on the grid 
$582\times 106 \times 64$ with $K=1$, not shown). Although the \kdtree\ with diffuser flow data gives better
agreement than the fine mesh,  the total shear stresses are
now even more over-predicted at $x=0.65$. Furthermore, the velocity profiles at $x=0.65$ exhibit
 small oscillations which probably stem from
a combination of low turbulent viscosity (the modeled shear stress is small where the oscillations appear, see Fig.~\ref{hump-uv-ni291}),
the central differencing  scheme and a coarse grid.

It is interesting to find  the location of the URANS/LES interface.
It is defined as the cell where $f_d$ falls below one, see Eq.~\ref{f_e}.
Figures~\ref{N-interface-diff} and ~\ref{N-interface-hump}  show how many cells are located in 
the URANS region.  Figures~\ref{yplus-interface-diff} and ~\ref{yplus-interface-hump} present the $y^+$ values of the
URANS/LES interface and it is seen that it is located in the range $0.06 \le y^+ \le 265$ 
and $1 \le y^+ \le 1\, 685$ for the diffuser flow and hump flow, respectively.
The $y^+$ values of the wall-adjacent cells are also presented in Figs.~\ref{yplus-interface-diff} and \ref{yplus-interface-hump}.
They are in the range $0.06 \le y^+ \le 35$ and $0.25 \le y^+ \le 50$ for the diffuser flow and hump flow, respectively.

\begin{figure}
\centering
\begin{subfigure}[t]{0.5\textwidth}
\centering\captionsetup{width=.8\linewidth}%
\centering
\includegraphics[scale=0.28,clip=]{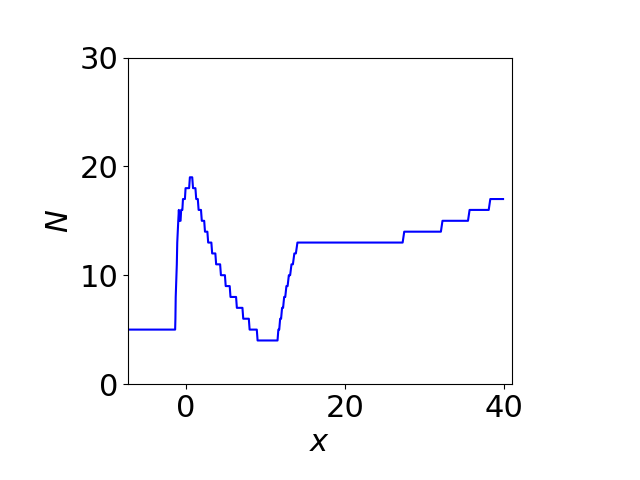}
\caption{Diffuser flow. Number of cells in URANS region}
\label{N-interface-diff}
\end{subfigure}%
\begin{subfigure}[t]{0.5\textwidth}
\centering\captionsetup{width=.8\linewidth}%
\centering
\includegraphics[scale=0.28,clip=]{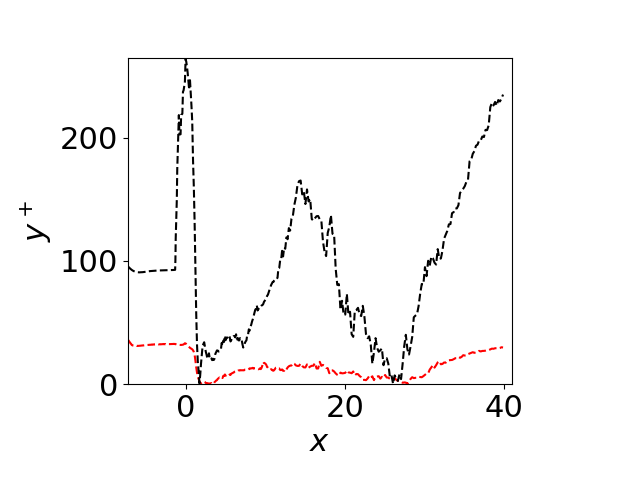}
\caption{Diffuser flow. \dashedline: $y^+$ of wall-adjacent cells. \dashedblackline: $y^+$ of URANS-LES interface.}
\label{yplus-interface-diff}
\end{subfigure}
\begin{subfigure}[t]{0.5\textwidth}
\centering\captionsetup{width=.8\linewidth}%
\centering
\includegraphics[scale=0.28,clip=]{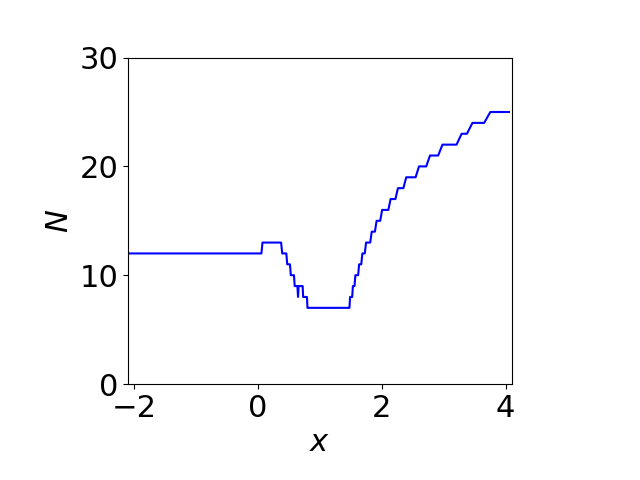}
\caption{Hump flow. Number of cells in URANS region}
\label{N-interface-hump}
\end{subfigure}%
\begin{subfigure}[t]{0.5\textwidth}
\centering\captionsetup{width=.8\linewidth}%
\centering
\includegraphics[scale=0.28,clip=]{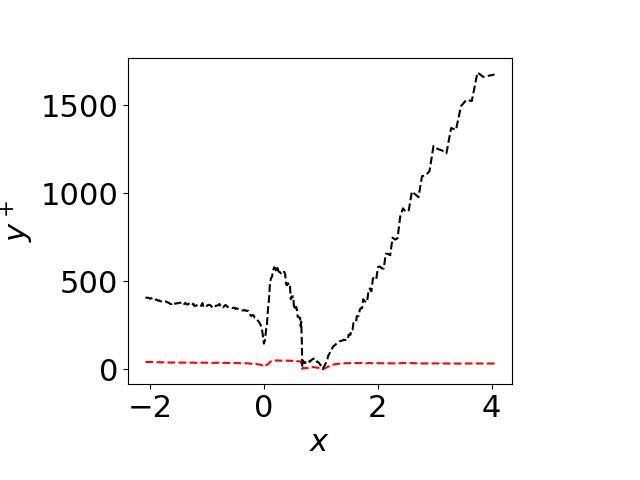}
\caption{Hump flow.\dashedline: $y^+$ of wall-adjacent cells. \dashedblackline: $y^+$ of URANS-LES interface.}
\label{yplus-interface-hump}
\end{subfigure}%
\hfill
\caption{Location of URANS-LES interface. Diffuser flow, $\a=15^o$. $350\times 70\times 48$ and Hump flow, $291\times 106\times 32$.}
\label{N-interface}
\end{figure}

In the wall function simulations presented so far, we have used the new wall function grids, see Fig.~\ref{new-wallf-grid}.
In the next case, we will use a standard wall function grid, see Fig.~\ref{wallf-grid}. The wall-adjacent cells ($j=0$) have the same size
as in the new wall function grid, and further away from the wall ($j>0$)  a geometrical stretching of $1.04$ is used
but with a limit $\Delta y < \Delta y_{max} \equiv  0.014$ ($\Delta y_{max}=0.015$
in the new wall function grid).
Figure~\ref{hump-flow-ni291-nj80} presents the results and it can be seen that the agreement with experiments is much worse than with the
new wall function grid.

\begin{figure}
\centering
\begin{subfigure}[t]{0.33\textwidth}
\centering
\includegraphics[scale=0.28,clip=]{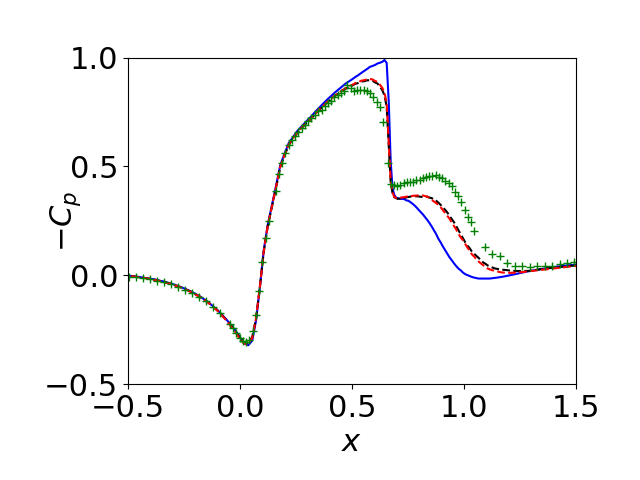}
\caption{Pressure coefficient.}
\end{subfigure}%
\hfill
\begin{subfigure}[t]{0.33\textwidth}
\centering
\includegraphics[scale=0.28,clip=]{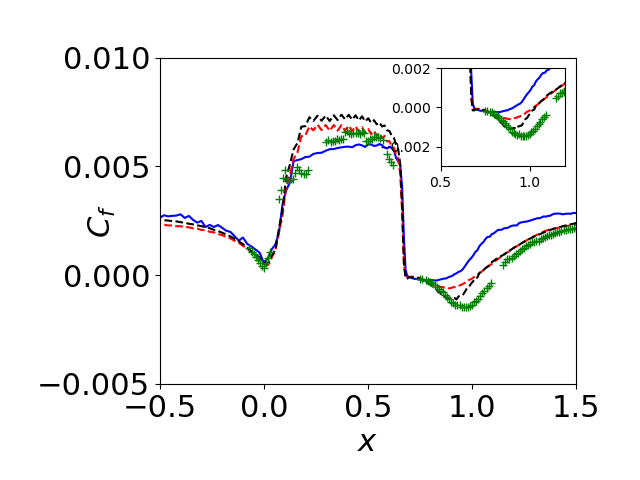}
\caption{Friction coefficient.}
\end{subfigure}%
\hfill
\begin{subfigure}[t]{0.33\textwidth}
\centering
\includegraphics[scale=0.28,clip=]{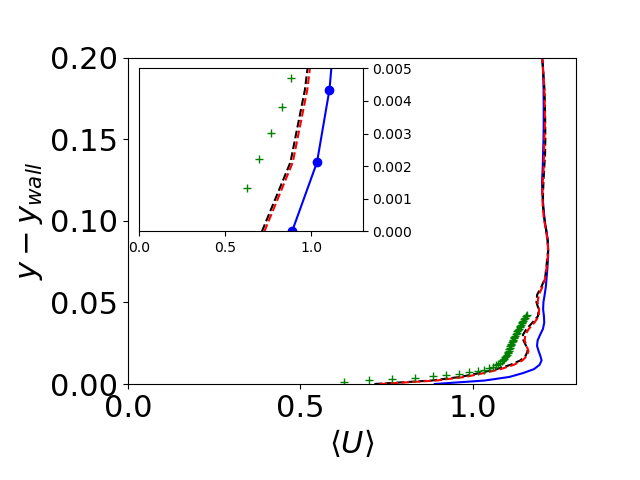}
\caption{$x=0.65$.}
\end{subfigure}
\begin{subfigure}[t]{0.33\textwidth}
\centering
\includegraphics[scale=0.28,clip=]{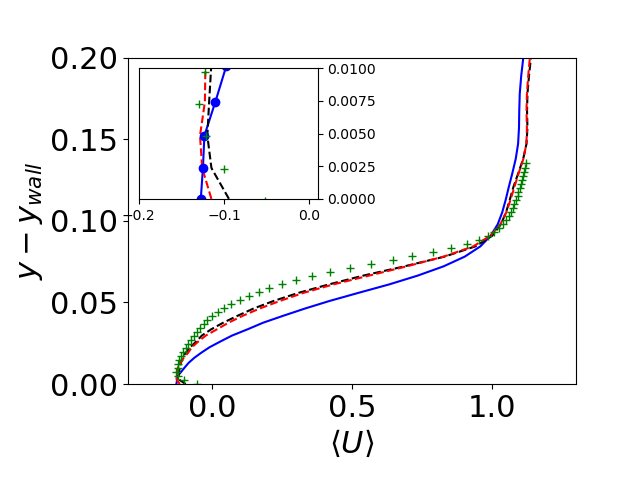}
\caption{$x=0.80$.}
\end{subfigure}%
\begin{subfigure}[t]{0.33\textwidth}
\centering
\includegraphics[scale=0.28,clip=]{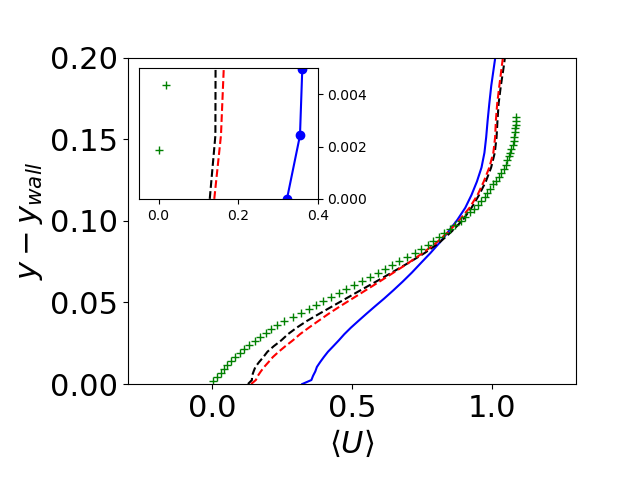}
\caption{$x=1.10$.}
\end{subfigure}%
\begin{subfigure}[t]{0.33\textwidth}
\centering
\includegraphics[scale=0.28,clip=]{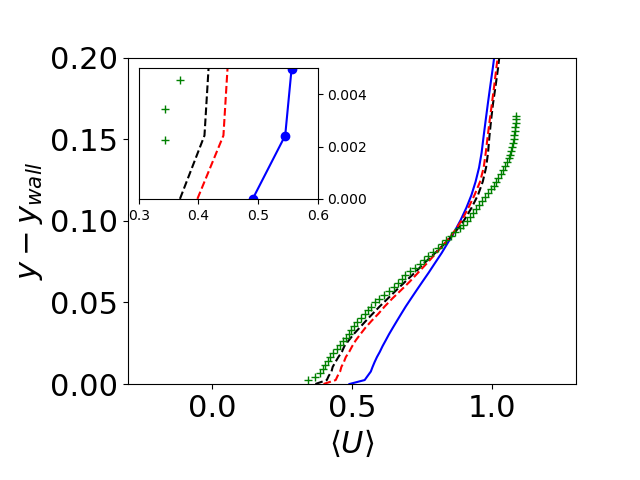}
\caption{Velocity at $x=1.30$.}
\end{subfigure}%
\caption{Hump flow. $(N_x \times N_z) = (291\times 32)$. $N_y = 80$. Standard wall-function mesh, see Fig.~\ref{wallf-grid}.
\solidline: \kdtree, hump flow data;  \dashedblackline: \kdtree, diffuser flow data;
  \dashedline: \wreich; \textcolor{blue}{$\bullet$}: location of cell centers;
\textcolor{green}{$+$}: experiments~\citep{greenblatt:04}.}
\label{hump-flow-ni291-nj80}
\end{figure}

\subsection{Flat-plate boundary layer flow}

\begin{figure}
\centering
\begin{subfigure}[t]{0.33\textwidth}
\centering
\includegraphics[scale=0.28,clip=]{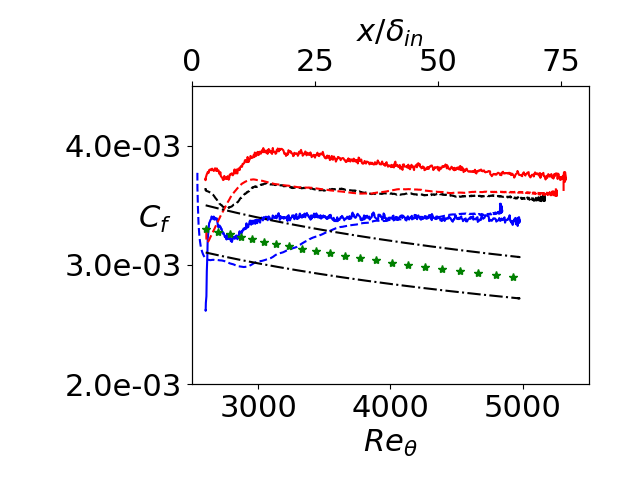}
\caption{Friction coefficient.}
\end{subfigure}%
\hfill
\begin{subfigure}[t]{0.33\textwidth}
\centering
\includegraphics[scale=0.28,clip=]{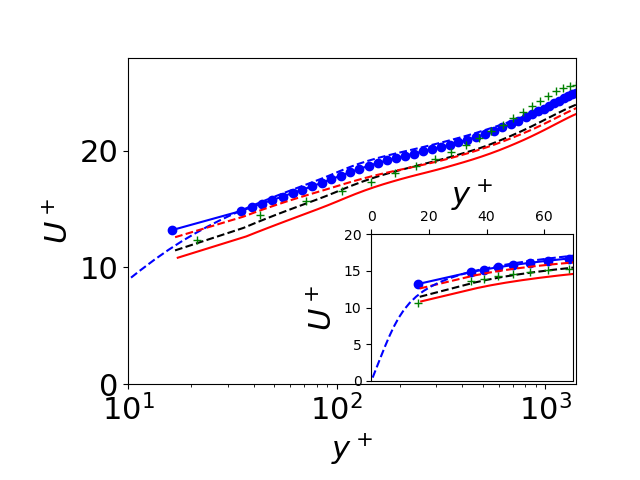}
\caption{Mean velocity.}
\end{subfigure}%
\hfill
\begin{subfigure}[t]{0.33\textwidth}
\centering
\includegraphics[scale=0.28,clip=]{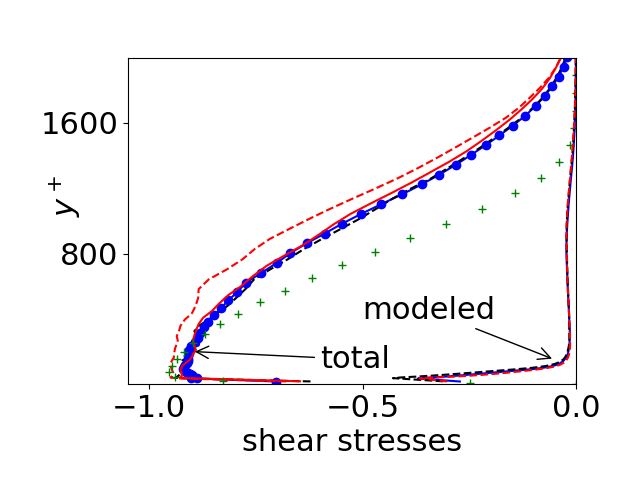}
\caption{Shear stresses.}
\label{bound-flow-uv}
\end{subfigure}
\caption{Boundary layer  flow. $u_\tau$ is computed by using $U^+$ and $y^+$ at the $3^{rd}$ cell. 
Velocity and  shear stresses are shown at $Re_\theta = 4\, 000$.
\solidline: \kdtree,  hump flow data
\dashedblackline: \kdtree, diffuser flow data;
\solidredline: \kdtree, diffuser flow data, $K=1$;
\dashedline: \wreich; \textcolor{blue}{$\bullet$}: location of cell centers; 
\dashedblueline: WR-IDDES;
\textcolor{green}{$\ast$}: $C_f = 
2(1/0.384 \ln (Re_\theta)+4.127)^{-2}$; \dashdottedline: $\pm 6\%$; 
\textcolor{green}{$+$}: DNS~\citep{sillero:13} (every $10^{th}$ cell).}
\label{bound-flow-3rd}
\end{figure}

The next test case is a flat-plate boundary layer.
It is a classic test case for verifying RANS turbulence models, for example the Reynolds stress model \citep{lrr:75} and
the SST model~\cite{menter:94}. It has also been used for evaluation hybrid LES/RANS models~\citep{deck:14} and 
embedded RANS-LES~\citep{shur:etal:11,davidson:15a}.
The Reynolds number at the inlet is $Re_{\theta}=2\, 550$.
The inlet and outlet 
boundary conditions are implemented in the same way as in the hump flow WR-IDDES simulations in 
 Section~\ref{database} and Fig.~\ref{inlet-bc}  (the mean profiles are taken 
from a pre-cursor 2D RANS at $Re_{\theta}=2\, 550$).
Periodic boundary conditions are set in the spanwise direction.
The grid has $550 \times 85 \times 64$ cells ($x$, $y$, $z$).
All length are made non-dimensional by the inlet boundary-layer thickness,  $\delta_{in}$ (unless viscous units are used).
The domain size is 
$7.9\times 5.7 \times 4.0$.
The first cell is obtained by merging $17$ cells in the WR-IDDES grid.
From the second wall-adjacent cell, the grid is stretched by $10\%$ for
$y < 0.64$ and $y> 2.5$ but
$\Delta y$ is not allowed to exceed $0.125$.
In the streamwise direction 
$\Delta x_{in}=0.1$ 
and a $0.1\%$ stretching is used
and the wall-adjacent cell center at the inlet is located at $y^+ =14$.

Figure~\ref{bound-flow-3rd} presents the skin friction, the velocity profiles and the turbulent shear stresses.
It is found that the \kdtree\ using hump flow data performs fairly well; it over-predicts $C_f$ by $12\%$ at $Re_\theta = 4\,000$ 
 (similar to the WR-IDDES and
better than the \wreich)
whereas the \kdtree\ using diffuser flow 
over-predicts the skin friction \kdtree\ by $18\%$.
The skin friction predicted with the \kdtree\ using diffuser flow data with $K=1$ is also shown in Fig.~\ref{bound-flow-3rd} and it is 
seen that $C_f$  is over-predicted by $25\%$. At $Re_\theta = 4\, 500$, for example, the predicted skin friction is $0.0038$, $0.0037$,
$0.0036$ and $0.0035$ for $K=1$, $K=2$, $K=5$ and $K=10$, respectively.
All three wall functions predict fairly good  velocities and shear stress profiles.

\begin{figure}
\centering
\begin{subfigure}[t]{0.33\textwidth}
\centering
\includegraphics[scale=0.28,clip=]{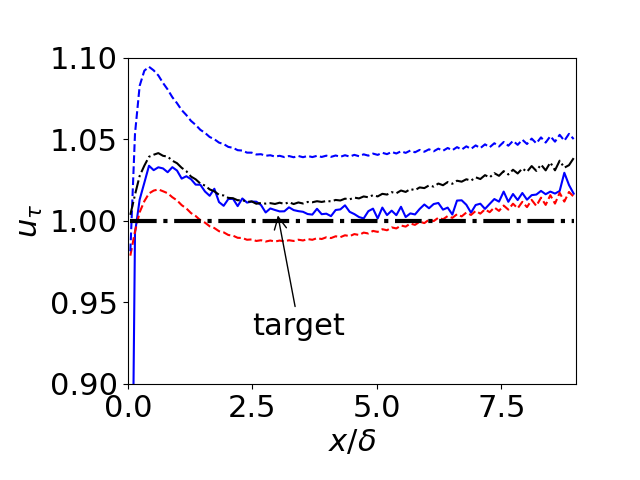}
\caption{Friction velocity.}
\label{channel-flow-ustar}
\end{subfigure}%
\hfill
\begin{subfigure}[t]{0.33\textwidth}
\centering
\includegraphics[scale=0.28,clip=]{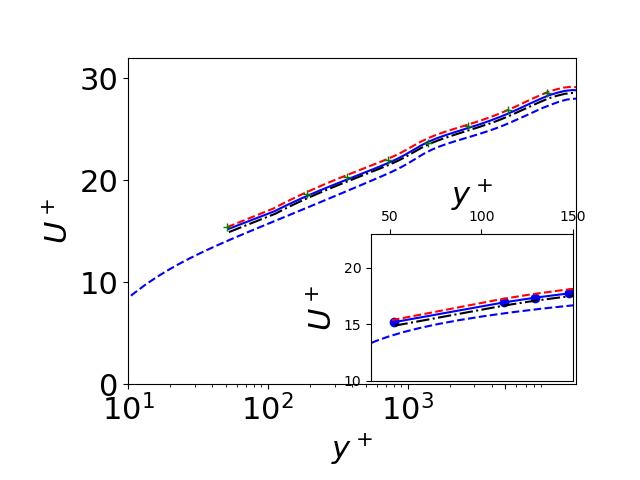}
\caption{Mean velocity.}
\end{subfigure}%
\hfill
\begin{subfigure}[t]{0.33\textwidth}
\centering
\includegraphics[scale=0.28,clip=]{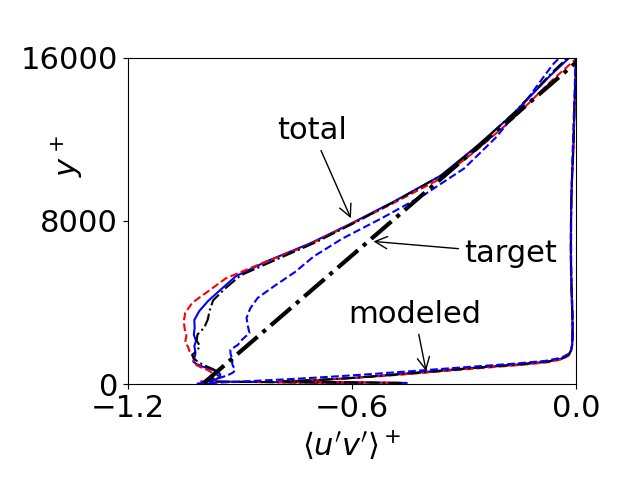}
\caption{Shear stress.}
\label{channel-flow-uv}
\end{subfigure}
\caption{Channel flow. $Re_\tau = 16\,000$.
Velocity and shear stress are shown at $x/\delta = 6$.
\solidline: \kdtree, hump flow data
\dashedblackline: \kdtree, diffuser flow data;
\dashedblueline: WR-IDDES, see Fig.~\ref{lowre-grid};
\solidredline:  \kdtree, hump flow data, $K=1$;
  \dashedline: \wreich; \textcolor{blue}{$\bullet$}: location of cell centers;
\textcolor{green}{$+$}: \reich, Eq.~\ref{reich}.} 
\label{channel-flow}
\end{figure}

\subsection{Channel flow}

The last test case is channel flow. 
 This test case is often used for evaluating inlet boundary 
conditions in LES~\citep{benhamadouche:01,mathey:06}  and hybrid RANS/LES~\citep{davidson:billson:05} as well as embedded 
RANS/LES~\citep{holgate:19}.
The domain size is $9\delta \times 2 \delta \times 1.6\delta$.
The Reynolds number at the inlet is $Re_\tau =16\, 000$ based on the friction velocity and the half-channel width, $\delta$. 
All length
are scaled with $\delta$ (except $y^+$). The inlet is located at $x=0$ and the outlet is at $x=9$.
The inlet and outlet 
boundary conditions are implemented in the same way as in the hump flow WR-IDDES simulations in 
Section~\ref{database} and Fig.~\ref{inlet-bc} (the mean profiles are taken 
from a pre-cursor 1D RANS at $Re_{\tau}=16\, 000$).
Periodic boundary conditions are set in the spanwise direction. The walls are located at $y=0$ and $y=2$.
The grid has  $96 \times 77 \times 32$ cells ($x$, $y$, $z$). Periodic boundary conditions are used for all variables in the spanwise ($z$)
direction.
The under-lying WR-grid is stretched by $14\%$ from the walls located at $y=0$ and $y=2$. A wall-adjacent cell in the 
wall-function grid is formed by merging $23$ near-wall cells in the WR-grid. 
The wall-adjacent cell center at the inlet is located at $y^+=45$.

Figure~\ref{channel-flow} shows the predicted friction velocity, the mean velocity and the turbulent shear stresses.
The \wreich\ and the \kdtree\ using hump flow data
predict the skin friction very well. The \kdtree{} using diffuser flow data and \wreich\ 
predict a skin friction which is approximately $2\%$ and $4\%$ too large, respectively, which also is good. 
On the other hand, the WR-IDDES yields a skin friction
which is $5\%$ too high. The target $u_\tau=1$ is also shown in Fig.~\ref{channel-flow-ustar}. The reason that no simulation reaches
the target is that the time-averaged streamwise velocity does not perfectly match the inlet
RANS velocity.  All three wall functions give too large a total shear stress which shows that the flow is not fully developed at $x=6$
(Fig.~\ref{channel-flow-uv}).

\section{Conclusions}

A new wall function based on  binary search trees, using Python' \kdtree\, has been presented.
Two different target database have been used, namely the flow in a diffuser (opening angle, $\a=15^o$) and the flow over a hump.
Time-averaged $U^+$ and $y^+$ are used as target quantities.

The reported resolution requirement in the literature for LES/DES of a boundary layer
using wall functions varies between $Re_L$ and $Re_L^{1.14}$. In the present work it is shown that wall-resolved IDDES (wall-adjacent
cell at $y^+ < 1$) requires an additional number of cells which vary with the friction Reynolds number as $\ln(Re_\tau)$
and $0.2\ln(Re_\tau)$ for wall-normal grid stretching of $1.04$ and $1.15$, respectively.

Five test cases have been  used:  the diffuser flow,  $\a=15^o$ and $\a=10^o$, the hump flow, flat-plate boundary layer flow  and channel 
flow.
The new \kdtree\ wall function using the diffuser flow database is found to perform well, better than when using the hump flow database
and better than the \wreich. 
The reason why the diffuser 
flow database performs better is probably because this is a simpler flow which makes it easier for the \kdtree\
search routine to find suitable $y^+$ and $U^+$.
The  new \kdtree\ wall function performs better that WR-IDDES for both flat-plate boundary layer and channel flow.
For the hump flow,  however, both  \kdtree\ wall functions  perform slightly worse than WR-IDDES. 
For example, the suction pressure peak is $4\%$ too low
and the velocity in the attached boundary layer upstream the re-circulation region is approximately $2\%$ too low. This 
results in somewhat under-predicted velocities in the recovery region downstream the re-circulation region.

A new strategy is used for creating the wall-normal distribution of the grid which was found to be very beneficial. The suction peak
of the pressure coefficient in the hump flow using the standard wall function grid is over predicted by $18\%$ (Fig.~\ref{hump-flow-ni291-nj80})
whereas the peak is well predicted using the new grid strategy (Fig.~\ref{hump-flow-ni291}).

The \kdtree\ using hump flow data  over-predicts the skin friction of the boundary flow  by $12\%$ at  $Re_\theta = 4\,000$ (same as WR-IDDES).
Using diffuser flow data in \kdtree\ gives an over-prediction of 
$18\%$ (same as \wreich). However, for this flow the \kdtree\ using diffuser flow data is sensitive to how many nearest neighbors are used
in \kdtree. Using the nearest neighbor, $K=1$ (instead of the baseline value of $K=5$), gives a $C_f$ which is $25\%$ too large.

It should be mentioned that databases with 
instantaneous data were also investigated, see end of Section~\ref{ML:sec} (no results were presented).
It was found that the instantaneous databases gave slightly worse results than the time-averaged ones.

\subsection*{Acknowledgments}

This study was financed by
\emph{Vinnova, NFFP, Grant No. 2023-01569},
 \emph{Strategic research project on Chalmers on hydro- and aerodynamics}.
and \emph{Chalmers Transport Area of Advance, Grant No. C 2023-0125-19}.

\appendix

\newpage

\begin{figure}
\centering
\begin{subfigure}[t]{0.5\textwidth}
\centering
\includegraphics[scale=0.31,clip=]{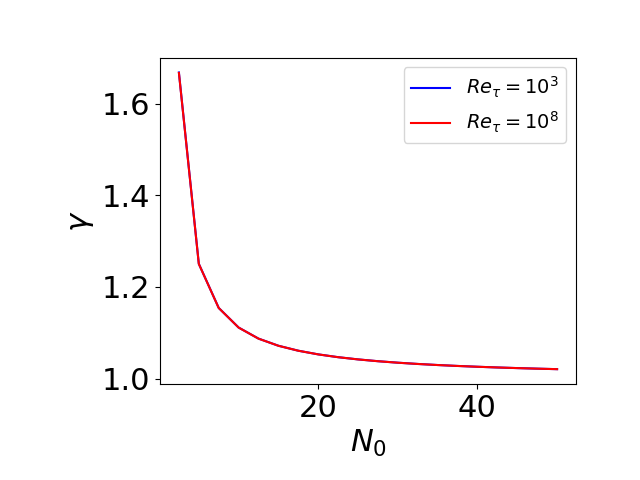}
\caption{Grid stretching versus $N_0$.}
\label{gamma}
\end{subfigure}%
\begin{subfigure}[t]{0.5\textwidth}
\centering
\includegraphics[scale=0.31,clip=]{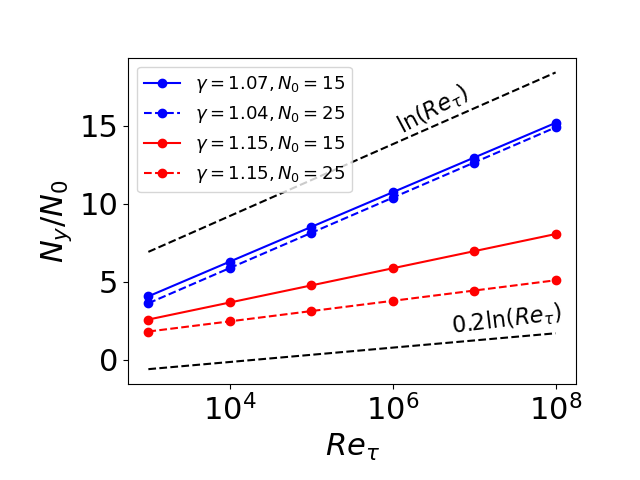}
\caption{Ratio of number of cells in stretched grids to that in WF grids.}
\label{ratio}
\end{subfigure}%
\caption{Comparison of WR-IDDES grids and wall-function grids.}
\end{figure}

\section{WR-IDDES grids}
\label{appendix-wallf}

We assume that the grid in the wall-normal direction is stretched by $\gamma$ the first $m$ cells, i.e.
\begin{eqnarray}
\label{series}
\Delta y + \gamma \Delta y + \gamma^2 \Delta y + \ldots + \gamma^m \Delta y
\end{eqnarray}
Setting $y^+ \equiv u_\tau \Delta y/\nu =1$ for the wall-adjacent 
cell and limit the largest wall-normal cell to $\delta/N_0$ ($\delta$ is the boundary-layer thickness) then
the last term in Eq.~\ref{series} reads
\begin{eqnarray}
\label{m}
\nu \gamma^m/u_\tau   = \frac{\delta}{N_0} \quad \Rightarrow \quad m \ln(\gamma) =  \ln(u_\tau\delta/(N_0\nu))   \quad \Rightarrow\nonumber \\
m =  \ln (Re_\tau/N_0)/ \ln(\gamma)
\end{eqnarray}
where $Re_\tau$ is the friction Reynolds number, $Re_\tau = u_\tau \delta/\nu$.
We define the region, $D_\gamma$,  in which the grid is stretched so that
\begin{eqnarray}
\label{D_gamma}
D_\gamma = \frac{\nu}{u_\tau} \left(1 +\gamma^1+\gamma^2 + \ldots +  \gamma^m\right) = \nonumber\\
= \frac{\nu}{u_\tau} \sum_{k=0}^m \gamma^k \quad \Rightarrow \quad 
D_\gamma = \frac{\nu}{u_\tau} \frac{1-\gamma^m}{1-\gamma}
\end{eqnarray}
Hence, the wall-normal grid of the boundary layer is formed by $m$ stretched cells which cover the region $D_\gamma$ and $N_y - m$ cells
of constant size $\delta/N_0$ which cover the region  $\delta - D_\gamma$ where $N_y$ is the total number of wall-normal cells.
For which grid
stretching, $\gamma$, does $D_\gamma$ extend all the way to $y=\delta$? By solving Eqs.~\ref{m} and \ref{D_gamma} numerically, 
we can find
the relation between $\gamma$ and $N_0$. We set $D_\gamma = \delta$ in Eq.~\ref{D_gamma} (recall that $\Delta y_{max} = \delta/N_0$, 
see Eq.~\ref{m})
\begin{eqnarray}
\label{solve-gamma}
 Re_\tau (1-\gamma) = 1- \gamma^m
\end{eqnarray}
and insert Eq.~\ref{m}
\begin{eqnarray}
\label{solve-gamma1}
Re_\tau (1-\gamma) = 1- \gamma^{\ln(Re_\tau/N_0)/\ln(\gamma)}
\end{eqnarray}
This equation is conveniently solved using the Newton-Raphson method, see 
Listing~\ref{solve-gamma-code} in Appendix~\ref{appendix-wallf-python}.

Figure~\ref{gamma} shows the solution of Eq.~\ref{solve-gamma1} for different Reynolds numbers. As can be seen, $\gamma$ versus $N_0$ is
virtually independent of Reynolds number. This can be understood by looking at Eq.~\ref{solve-gamma}.
 A large change in $Re_\tau$ gives only a tiny change in $m$ (recall that $50 \lesssim m \lesssim 200$). 
The ratio $N_y/N_0$ versus Reynolds number is in Fig.\ref{ratio} shown
for two combinations of ($\gamma$, $N_0$) that satisfy Eq.~\ref{solve-gamma1}, i.e. 
$(1.07,15)$ and $(1.04, 25)$ which are shown by  blue lines; recall that when  Eq.~\ref{solve-gamma1} is satisfied 
$\Delta y < \delta/N_0$ for all cells and $m=N_y$. However,
for the other two grids in Fig.~\ref{ratio} (red lines), there are a number of cells of constant size, $\delta/N_0$,
 outside the region with stretched cells. At $Re_\tau=10^8$, for
example, $(N_y,m)$ are equal
to $(120,113)$ and  $(127, 109)$ for $N_0=15$ and  $N_0=25$, respectively.
Black dashed lines in Fig.~\ref{ratio} show upper ($\ln(Re_\tau)$)
and lower trend ($0.2\ln(Re_\tau)$) versus Reynolds number.

\newpage

\section{Python script for comparing stretched and wall-function grids}

\label{appendix-wallf-python}


\begin{lstlisting}[caption={Python script for solving Eq.~\ref{solve-gamma1} for different $N_0$ and $Re_\tau$},captionpos=b,label={solve-gamma-code}]
import numpy as np
import sys
import matplotlib.pyplot as plt
from scipy.optimize import fsolve,root,newton

plt.rcParams.update({'font.size': 22})
plt.close('all')

plt.interactive(True)

def solve_gamma(gam,delta,dx):
   return  (1-gam**(np.log(Re/N_0)/np.log(gam)+1))\
             - Re*(1-gam)

# solve for n_N0 different N_0
n_N0 = 20
# solve for n_re different re
n_re = 6
# set lowest Reynolds number
Re = 1e3
gamma_vector = np.zeros((n_re,n_N0))
Re_vector = np.zeros((n_re))
N_0_vector = np.zeros((n_N0))
m_vector = np.zeros((n_re,n_N0))
dy_vector = np.zeros((n_re,n_N0))

# loop over Re
for r in range(0,n_re):
# first N_0
   N_0 = 2.5
   Re_vector[r] = Re
# loop over N_0
   for n in range(0,n_N0):
      
# intial gamma value into the solver
      gamma = 1.01

# call the Newton-Raphson solver
      gamma = newton(solve_gamma,x0=gamma,\
              args=(Re,N_0))

      N_0_vector[n] = N_0

# compute m
      m_vector[r,n] = np.log(Re/N_0)/np.log(gamma)

# largest dy
      dy_vector[r,n] = gamma**m_vector[r,n]

      gamma_vector[r,n] = gamma

# next N_0
      N_0 = N_0 +2.5

# next Re
   Re = Re*10

######################## plot gamma vs N_0
fig1,ax1 = plt.subplots()
plt.subplots_adjust(left=0.25,bottom=0.20)
# smallest Reynolds number 
plt.plot(N_0_vector,gamma_vector[0,:],'b-')
# largest Reynolds number 
plt.plot(N_0_vector,gamma_vector[-1,:],'r-')
plt.xlabel(r"$N_0$")
plt.ylabel(r"$\gamma$")
plt.savefig('gamma-vs-N0-all-N0-and-re.png')
\end{lstlisting}

\newpage

\section{Python script for \kdtree\ wall function}
\label{appendix-kdtree}

\begin{lstlisting}[caption={The \texttt{fix\_k} module which includes the ML (\kdtree) wall function. Line number are shown to the right.}
,captionpos=b,label={ML-code}]
def fix_k():                                                   
    from sklearn.preprocessing import MinMaxScaler             
    from scipy.spatial import KDTree                           
    global X, tree, yplus_target                               
                                                               
    if iter == 0 and itstep == 0:                              
#at t start-up: load target data                                
       data = xp.loadtxt('x-yplus-uplus-diffuser.txt')      
       x_target = data[:,0]                                 
       yplus_target = abs(data[:,1])                        
       uplus_target = data[:,2]                             
                                                               
       uplus_target = uplus_target.reshape(-1,1)               
       yplus_target = yplus_target.reshape(-1,1)               
# use MinMax scaler                                            
       scaler_yplus = MinMaxScaler()                           
       scaler_uplus = MinMaxScaler()                           
                                                               
# store and scale target in X                                  
       X=xp.zeros((len(yplus_target),2))                       
       X[:,0] = scaler_uplus.fit_transform(\
                uplus_target)[:,0]  
       X[:,1] = scaler_yplus.fit_transform(\
                yplus_target)[:,0]  
# build the tree                                               
       tree = KDTree(X)                                        
# the code below is executed every CFD iteration               
# take old ustar from previous iteration/time step             
    ustar=cmu**0.25*k3d[:,0,:]**0.5                            
# create a 2D arraay with wall-parallel velocity               
    j_wall = 0
    u2d_wall=u3d[:,j_wall,:]                                   
# cell-center wall distance                                    
    dy=dist3d[:,j_wall,:]                                      
# compute yplus and uplus                                      
    yplus_south = ustar*dy/viscos                              
    uplus_south = u2d_wall/ustar                               
                                                               
    yplus = yplus_south.reshape(-1,1)                          
    uplus = uplus_south.reshape(-1,1)                          
                                                               
# store and scale yplus and uplus from CFD in x                
    x=xp.zeros((len(uplus),2))                                 
    x[:,0] = scaler_uplus.transform(uplus)[:,0]                
    x[:,1] = scaler_yplus.transform(yplus)[:,0]                
                                                               
# find one (K=1) nearest neighbor at distance ds               
    K=1                                                        
    ds, inds =  tree.query(x_np, K)                            
                                                               
# set yplus and reshape                                        
    yplus_kdtree = yplus_target[inds,0]                        
    yplus_predict = xp.reshape(yplus_kdtree,(ni,nk))           
# compute u_tau                                                
    ustar=yplus_predict*viscos/dy                              
                                                               
# compute k (turb. kinetic energy)                             
    kwall=cmu**(-0.5)*ustar**2                                 
                                                               
# fix k at wall-adjacent cells                                 
    aw3d[:,0,:]=0                                              
    ae3d[:,0,:]=0                                              
    as3d[:,0,:]=0                                              
    an3d[:,0,:]=0                                              
    al3d[:,0,:]=0                                              
    ah3d[:,0,:]=0                                              
    ap_max=xp.max(ap3d)                                        
    ap3d[:,0,:]=ap_max                                         
    su3d[:,0,:]=ap_max*kwall                                   

    return aw3d,ae3d,as3d,an3d,al3d,ah3d,ap3d,su3d,sp3d

\end{lstlisting}

\newpage


\clearpage

\end{document}